\documentclass[aps,pra,twocolumn,superscriptaddress,amsmath,amssymb]{revtex4-2}
\usepackage[colorlinks,pdfusetitle,urlcolor=blue,citecolor=blue,linkcolor=blue,bookmarksnumbered,plainpages=false]{hyperref}
\usepackage{graphicx}
\usepackage{bm}

\renewcommand{\vec}{\mathbf}

\begin{document}
\title{Hidden Momenta due to Magneto-Electric Coupling in Non-Reciprocal Media}

\author{William S. McGregor}
\affiliation{School of Physics \& Astronomy, University of Glasgow, Glasgow G12 8QQ, United Kingdom}
\author{Niclas Westerberg}
\affiliation{SUPA, Institute of Photonics and Quantum Sciences, Heriot-Watt University, Edinburgh EH14 4AS, United Kingdom}
\affiliation{School of Physics \& Astronomy, University of Glasgow, Glasgow G12 8QQ, United Kingdom}
\author{Robert Bennett}
\affiliation{School of Physics \& Astronomy, University of Glasgow, Glasgow G12 8QQ, United Kingdom}
\date{\today}

\begin{abstract} 
From a canonical quantisation approach of macroscopic quantum electrodynamics, we derive the interaction Hamiltonian due to a dipole emitter interacting with a time-symmetry-broken topological insulator (TSB-TI), described by an axionic coupling Lagrangian. This provides an additional interaction term dependent on the magneto-electric coupling strength inherent to the TSB-TI which will modify measurements on related observables. The spontaneous decay rate is calculated in the Born-Markov approximation for a general electromagnetic Green's tensor, and then specifically for the case of a planar parallel-plate cavity. 
\end{abstract}

\maketitle

\section{Introduction} \label{S1}

In systems with electromagnetic interactions, the momentum conjugate to a coordinate need not coincide with the corresponding mechanical momentum. The difference between these quantities is often referred to as \emph{hidden} momentum: that stored in the coupled field–matter degrees of freedom rather than in the mechanical motion of the system \cite{shockleyTrySimplestCases1967}. A familiar example arises from the Röntgen interaction \cite{rontgen1888ueber}, in which the motion of an electric dipole $\mathbf{d}$ through electromagnetic fields induces an effective coupling to the magnetic field, equivalently making the moving dipole behave as though it carries a magnetic moment. This can be applied in a variety of scenarios such as investigating how the force on an emitter is modified due to the additional interaction term ~\cite{sonnleitner2017rontgen}. Such effects are not captured by a purely electrostatic interaction with the electric field $\mathbf{E}$ of the form $-\mathbf{d}\cdot\mathbf{E}$, as it omits the velocity-dependent and magneto-electric terms required by a consistent dynamical description -- something that is important to, for instance, find the correct form of the spontaneous decay rate of the moving atom \cite{HiddenMomentum_movingAtom}. More generally, whenever electric and magnetic responses are coupled, the resulting distinction between canonical and mechanical momentum can influence measurable quantities. Indeed, this distinction was paramount for resolving the Abraham-Minkowski controversy on the correct form of the momentum of light in optical media \cite{HiddenMomentum_AMdilemma_Steve}, including when magnetic charges are included \cite{HiddenMomentum_AMdebate_Saldanha}. It also enters in our understanding of momentum exchange between electrons and current loops \cite{HiddenMomentum_DiracElectron}, and in the differences between the electromagnetic force distributions possible \cite{HiddenMomentum_ForceDistri_Wright}. This is furthermore closely connected to the Aharonov-Bohm \cite{ABeffect} and Aharonov-Casher \cite{ACeffect} effects, i.e. the quantum mechanical phases accumulated by moving objects near electromagnetic fields and potentials, and particularly how the latter is derived for neutral systems \cite{HiddenMomentum_ACeffect_Horsley}. The distinction is even important for hydrogen atoms moving in static electric fields \cite{HiddenMomentum_Hydrogen}, the static electromagnetic fields themselves \cite{HiddenMomentum_StaticCharges}, and it played a crucial role in resolving the apparent conflict between diamagnetism and causality \cite{HiddenMomentum_Diamagnetism_0, HiddenMomentum_Diamagnetism_1}. It is therefore natural to ask whether hidden-momentum effects play a role in materials with intrinsic magneto-electric coupling.

A particularly relevant example is a time-symmetry-broken topological insulator (TSB-TI). In such a material, the breaking of time-reversal symmetry permits a non-reciprocal electromagnetic response described, at the level of the effective Lagrangian, by an axion-like coupling proportional to $\mathbf{E}\cdot\mathbf{B}$ \cite{wilczek1987two}. Equivalently, the constitutive relations no longer take the purely diagonal form in which $\mathbf{D}$ is proportional only to $\mathbf{E}$ and $\mathbf{H}$ only to $\mathbf{B}$. Instead, electric and magnetic fields enter each other’s response equations through magneto-electric mixing terms. This makes a time-symmetry-broken topological insulator a natural setting for investigating how hidden momentum emerges from a fully consistent action principle, and how it may enter observable electromagnetic response.

Typically, calculations of observables in non-reciprocal media proceed in terms of the usual electric dipole coupling, with the non-reciprocal nature of the medium entering through the specific form of $\mathbf{E}$ relevant to the boundary conditions that the non-reciprocal medium imposes \cite{buhmann2012macroscopic, fuchs2017casimir}. Here we will show that additional contributions arise from a careful account of the Lagrangian formulation of the problem, and that these have quantifiable effects when the quantized version of the theory is used to calculate the spontaneous decay rate of an atom.

The structure of this paper is as follows. In Section \ref{S2} the Lagrangian description for the system described in Ref.~\cite{crosse2015electromagnetic} will be employed with the help of the macroscopic QED \cite{grunerGreenfunctionApproachRadiationfield1996,buhmann2013dispersion,buhmann2013dispersionII, MQED_Rob_Nic} canonical quantisation framework derived by Philbin~\cite{philbin2010canonical}, in which the canonical momenta and interaction Hamiltonian for a system of a dipolar emitter interacting with a TSB-TI will be derived. Appendix~\ref{AppendixA} will show that the results and derivations in Section \ref{S2} are consistent with the phenomenological approach utilised in Ref.~\cite{crosse2015electromagnetic} and Ref.~\cite{fuchs2017casimir}. In Section \ref{S4}, as a demonstration of how these hidden momenta impact observables, the spontaneous decay rate will be calculated for a general Green's tensor, and then in Section \ref{S5} applied specifically to the case of a planar cavity non-reciprocal Green's tensor in order to demonstrate the quantifiable differences between the usual calculation of the decay rate and the full canonical quantisation treatment.

\section{Non-Reciprocal Lagrangian \& Hamiltonian} \label{S2}

The axionic Lagrangian describing the magneto-electric coupling and lack of time-reversal symmetry in non-reciprocal media for TSB-TIs is (see, e.g. Ref~\cite{fuchs2017casimir});
\begin{equation} \label{L_NR}
    \mathcal{L}_\mathrm{NR} = \frac{\alpha \theta(\vec{r})}{2 \pi Z_0}\biggl(\vec{E}(\vec{r},t) \cdot \vec{B}(\vec{r},t) + \vec{B}(\vec{r},t) \cdot \vec{E}(\vec{r},t)\biggl),
\end{equation}
where $Z_0 = \mu_0 c$ is the vacuum impedance, $\alpha$ is the fine-structure constant, $\vec{E}$ and $\vec{B}$ are the usual electric and magnetic fields, and $\theta(\vec{r})$ is the axionic coupling factor, which is a real, dimensionless number that can be considered a measure of the non-reciprocity of the medium in consideration. This type of interaction has been employed in several works to describe interactions with non-reciprocal media~\cite{chang2009optical, crosse2015electromagnetic, fuchs2017casimir, franca2019reversed, franca2024vavilov}. Note that Eq.~\eqref{L_NR} has been written in an hermitian form which will prove useful during the quantisation step.

An additional interaction Lagrangian, $\mathcal{L}_d$, describes the interaction with the emitter at position $\vec{r}_\mathrm{A}$ in the electric-dipole approximation.
\begin{equation} \label{dipoleint}
\mathcal{L}_d = -\rho_d (\vec{r},t) \phi(\vec{r},t) + \vec{J}_d(\vec{r},t) \cdot \vec{A}(\vec{r},t),
\end{equation}
where,
\begin{align}
&\rho_d = -\vec{d}(\vec{r}_\mathrm{A},t) \cdot \nabla \delta (\vec{r}-\vec{r}_\mathrm{A}), \label{rho}\\
&\vec{J}_d = \dot{\vec{d}}(\vec{r}_\mathrm{A},t) \delta (\vec{r}-\vec{r}_\mathrm{A}), \label{J}
\end{align}
respectively represent the charge and current densities of the dipole moment, $\vec{d}$, while $\phi$ and $\vec{A}$ represent the scalar and vector potentials of the electromagnetic field. The multipolar coupling scheme will be considered, which corresponds to a transformation of the action pertaining to Eq.~\eqref{dipoleint}, by first expanding each term via the product rule.
\begin{align}\label{Sd}
 S_d = \int d^4x&\biggl(\nabla \cdot \left[\vec{d}(\vec{r}_\mathrm{A},t) \phi(\vec{r},t) \delta(\vec{r}-\vec{r}_\mathrm{A})\right] \notag \\
& \quad -\vec{d}(\vec{r}_\mathrm{A},t) \cdot \nabla \phi(\vec{r},t) \delta(\vec{r}-\vec{r}_\mathrm{A})\notag \\
&\quad \quad  + \delta(\vec{r}-\vec{r}_\mathrm{A}) \biggl[ \frac{d}{dt}(\vec{d}(\vec{r}_\mathrm{A},t) \cdot \vec{A}(\vec{r},t)) \notag \\
&\quad \quad \quad  - \vec{d}(\vec{r}_\mathrm{A},t) \cdot \dot{\vec{A}}(\vec{r},t) \biggl]\biggl).
\end{align}
The first term in Eq.~\eqref{Sd} is zero due to a vanishing surface integral from the divergence theorem, and the third term is a total time derivative, so will not affect the dynamics. Therefore, the dipole interaction Lagrangian can be written in the multipolar coupling scheme as
\begin{equation} \label{L_dipole}
\mathcal{L}_\mathrm{d} = -\vec{d}_{\rho}(\vec{r},t) \cdot \left[ \nabla \phi (\vec{r},t) + \dot{\vec{A}}(\vec{r},t) \right], 
\end{equation}
where $\vec{d}_{\rho}(\vec{r},t) = \vec{d}(\vec{r}_\mathrm{A},t) \delta(\vec{r} - \vec{r}_\mathrm{A})$ is the dipole moment density. From this point onward, the spatial- and time-dependence of the dynamical variables will be suppressed for ease of reading. Thus, the total interaction Lagrangian for the non-reciprocal magneto-electric coupling and dipole interaction can be written via Eqs.~\eqref{L_NR} and \eqref{L_dipole} as  
\begin{equation} \label{LT}
\mathcal{L}_\mathrm{int} = \mathcal{L}_\mathrm{NR} + \mathcal{L}_\mathrm{d}.
\end{equation}
This interaction Lagrangian forms the basis of our subsequent canonical quantisation of the electromagnetic field in the presence of absorbing, dispersive media. This was first carried out via the Huttner-Barnett model~\cite{huttner1992quantization} and was later generalised to include magnetic excitations by Philbin in the context of macroscopic QED~\cite{philbin2010canonical}. In the same spirit as~\cite{philbin2010canonical} and \cite{huttner1992quantization}, which consider the special case of reciprocal media, we develop an extension to non-reciprocal media by constructing the total Lagrangian,  $\mathcal{L}=\mathcal{L}_\mathrm{R} + \mathcal{L}_\mathrm{int}$, where, 
\begin{equation} \label{L_0}
\mathcal{L}_\mathrm{R} = \mathcal{L}_\mathrm{EM} + \mathcal{L}_\mathrm{X} + \mathcal{L}_\mathrm{Y} + \mathcal{L}_\mathrm{I},
\end{equation}
which is the usual Lagrangian structure for reciprocal media derived in Ref.~\cite{philbin2010canonical} with, 
\begin{align}
& \mathcal{L}_\mathrm{EM} = \frac{\varepsilon_0}{2}\vec{E} \cdot \vec{E} - \frac{1}{2 \mu_0}\vec{B} \cdot \vec{B}, \label{LEM} \\
& \mathcal{L}_\mathrm{X} = \frac{1}{2}\int_0^{\infty} d\omega \, \biggl(\dot{\vec{X}}_{\omega} \cdot \dot{\vec{X}}_{\omega} - \omega^2 \vec{X}_{\omega} \cdot \vec{X}_{\omega} \biggl), \label{LX}\\
& \mathcal{L}_\mathrm{Y} = \frac{1}{2}\int_0^{\infty} d\omega \, \biggl(\dot{\vec{Y}}_{\omega} \cdot \dot{\vec{Y}}_{\omega} - \omega^2 \vec{Y}_{\omega} \cdot \vec{Y}_{\omega} \biggl), \label{LY}\\
& \mathcal{L}_\mathrm{I} = \int_0^{\infty} d\omega \, \biggl(\alpha(\vec{r},\omega)\vec{X}_{\omega} \cdot \vec{E} + \beta(\vec{r},\omega) \vec{Y}_{\omega} \cdot \vec{B} \biggl). \label{LXYINT}
\end{align}
$\mathcal{L}_\mathrm{EM}$ is the usual Lagrangian for the electromagnetic field in vacuum. $\vec{X}_{\omega}$ and $\vec{Y}_{\omega}$ correspond to the oscillator fields of the electric and magnetic reservoirs respectively. These fields represent the matter degrees of freedom in the medium and the Lagrangian densities $\mathcal{L}_\mathrm{X}$ and $\mathcal{L}_\mathrm{Y}$ are modelled as harmonic oscillators to describe the typical driven-oscillation dynamics that incur when matter in a medium interacts with the electromagnetic field. Lastly, $\mathcal{L}_\mathrm{I}$ corresponds to the interaction of the electromagnetic field with the continuum of matter degrees of freedom through the absorptive nature of the medium. The electric and magnetic coupling factors for these interactions, $\alpha$ and $\beta$ are given by, 
\begin{align}
& \alpha(\vec{r},\omega) = \sqrt{\frac{2\varepsilon_0}{\pi}\omega \varepsilon_I(\vec{r},\omega)}, \label{alpha}\\
&\beta(\vec{r},\omega) = \sqrt{\frac{2}{\pi \mu_0}\omega \frac{\mu_I(\vec{r},\omega)}{\vert \mu(\vec{r},\omega)\vert^2}}. \label{beta}
\end{align}
where $\varepsilon_I$ and $\mu_I$ are the imaginary parts of the relative permittivity and permeability respectively of a generally absorbing, dispersive, inhomogeneous medium. These represent the interaction of the electromagnetic field with the matter degrees of freedom through absorption, and can be related to their real parts via the Kramers-Kronig relations, 
\begin{align}
& \varepsilon_R(\vec{r},\omega) = 1 + \frac{2}{\pi}\mathcal{P}\int_0^{\infty} d\omega'\, \frac{\omega' \varepsilon_I (\vec{r},\omega')}{\omega'^2 -\omega^2}, \label{KKR}\\
& \varepsilon_I(\vec{r},\omega) =  -\frac{2\omega}{\pi}\mathcal{P}\int_0^{\infty} d\omega'\, \frac{ \varepsilon_R (\vec{r},\omega') - 1}{\omega'^2 -\omega^2}, \label{KKI}
\end{align}
and similarly for $\mu(\vec{r},\omega)$.
\subsection{Equations of Motion} \label{S2a}

The total action functional of the system described so far can be represented by the four dynamical variables: $\vec{A}$, $\phi$, $\vec{X}_{\omega}$, and $\vec{Y}_{\omega}$. Variation of the action to produce the Euler-Lagrange equations of motion has already been performed for $\mathcal{L}_\mathrm{R}$ in Ref.~\cite{philbin2010canonical}. So, we derive only the contributions due to the additional Lagrangian term, $\mathcal{L}_\mathrm{int}$. The equations of motion for $\vec{X}_{\omega}$ and $\vec{Y}_{\omega}$ are unchanged relative to Ref.~\cite{philbin2010canonical} as they do not feature in $\mathcal{L}_\mathrm{int}$, however, the equations of motion for $\vec{A}$ and $\phi$;
\begin{align}
& \partial_t \left(\frac{\partial\mathcal{L}}{\partial\dot{\phi}}\right) + \nabla_{\beta}\left(\frac{\partial\mathcal{L}}{\partial\nabla_{\beta}\phi} \right) = \frac{\partial\mathcal{L}}{\partial\phi}, \label{ELphi}\\
& \partial_t \left(\frac{\partial\mathcal{L}}{\partial\dot{A_{\alpha}}}\right) + \nabla_{\beta}\left(\frac{\partial\mathcal{L}}{\partial\nabla_{\beta}A_{\alpha}} \right) = \frac{\partial\mathcal{L}}{\partial A_{\alpha}}, \label{ELA}
\end{align}
gain additional terms. The derivatives pertaining to $\mathcal{L}_\mathrm{int}$ are, 
\begin{align}
& \partial_t \left(\frac{\partial\mathcal{L}_\mathrm{int}}{\partial\dot{\phi}}\right) = 0, \label{phitD}\\
& \nabla_{\beta}\left(\frac{\partial\mathcal{L}_\mathrm{int}}{\partial\nabla_{\beta}\phi} \right) = -\frac{\alpha}{\pi}Z_0^{-1}\nabla_{\beta} \left(\theta B_{\beta} \right) - \nabla_{\beta}d_{\rho,\beta}, \label{phispaceD}\\
& \partial_t \left(\frac{\partial\mathcal{L}_\mathrm{int}}{\partial\dot{A_{\alpha}}}\right) = -\frac{\alpha}{\pi}Z_0^{-1}\partial_t \left(\theta B_{\alpha} \right) - \dot{d}_{\rho, \alpha}, \label{AtD}\\
& \nabla_{\beta}\left(\frac{\partial\mathcal{L}_\mathrm{int}}{\partial\nabla_{\beta}A_{\alpha}} \right) = -\frac{\alpha}{\pi}Z_0^{-1}\varepsilon^{\alpha \beta \gamma}\nabla_{\beta} \left(\theta E_{\gamma} \right). \label{AspaceD}
\end{align}
Combining these with the equations of motion derived for $\mathcal{L}_\mathrm{R}$ in~\cite{philbin2010canonical}, the following equations of motion are obtained for $\mathcal{L}$:
\begin{align}
\varepsilon_0 \nabla \cdot \textbf{E}  =& -\int_{0}^{\infty} d\omega \, \nabla\cdot \left[\alpha(\textbf{r}, \omega)\textbf{X}_{\omega} \right]\notag \\
& \qquad-\nabla \cdot \vec{d}_{\rho}  - \frac{\alpha}{\pi}Z_0^{-1}\nabla \theta \cdot \textbf{B}, \label{phiEOM}\\
\frac{1}{\mu_0} \nabla \times \textbf{B} =&\, \varepsilon_0 \dot{\textbf{E}} +\int_{0}^{\infty}\,d\omega \bigl( \alpha(\textbf{r}, \omega)\dot{\textbf{X}}_{\omega} \notag \\
& + \nabla \times \left[\beta(\textbf{r}, \omega)\textbf{Y}_{\omega} \right]\bigl) \notag \\
& \quad + \frac{\alpha}{\pi}Z_0^{-1}\nabla \theta \times \textbf{E} + \dot{\vec{d}}_{\rho}, \label{AEOM}\\
&\ddot{\textbf{X}}_{\omega} + \omega^2 \textbf{X}_{\omega} = \alpha(\textbf{r},\omega)\textbf{E}, \label{XEOM}\\
&\ddot{\textbf{Y}}_{\omega} + \omega^2 \textbf{Y}_{\omega} = \beta(\textbf{r},\omega)\textbf{B}. \label{YEOM}
\end{align}
For the reciprocal case with no free charges ($\theta=0$ and $\vec{d}_{\rho}=0$), Eqs.~\eqref{phiEOM}--\eqref{YEOM} resemble the usual Maxwell's equations in media~\cite{jackson2012classical} provided that the following identifications for the polarisation and magnetisation are made:
\begin{align}
& \vec{P} = \int_0^{\infty} d \omega \, \alpha(\vec{r},\omega) \vec{X}_{\omega}, \label{Polarisation}\\
& \vec{M} = \int_0^{\infty} d \omega \, \beta(\vec{r},\omega) \vec{Y}_{\omega}. \label{Magnetisation}
\end{align}
Since the polarisation and magnetisation depend purely on the matter degrees of freedom, Eqs.~\eqref{Polarisation} and \eqref{Magnetisation} still hold for the non-reciprocal case. This provides the first indication that in considering $\mathcal{L}_\mathrm{NR}$, it is the constitutive relations for the displacement field, $\vec{D}$, and the auxiliary field, $\vec{H}$, that will be modified due to this interaction.

\subsection{Canonical Momenta} \label{S2b}

Similarly to the equations of motion, the field momenta for $\vec{X}_{\omega}$ and $\vec{Y}_{\omega}$ will be unchanged relative to the reciprocal case in Ref.~\cite{philbin2010canonical}, and are simply given by:
\begin{align}
& \vec{\Pi}_{\vec{X}_{\omega}} = \frac{\partial \mathcal{L} }{\partial \vec{\dot{X}}_{\omega}} = \vec{\dot{X}}_{\omega}, \label{Xpi}\\
& \vec{\Pi}_{\vec{Y}_{\omega}} = \frac{\partial \mathcal{L} }{\partial \vec{\dot{Y}}_{\omega}} = \vec{\dot{Y}}_{\omega}. \label{Ypi}
\end{align}
Furthermore, since $\mathcal{L}$ does not feature $\dot{\phi}$ explicitly, then $\vec{\Pi}_{\phi}=0$. The more interesting and consequential canonical momentum to the Hamiltonian structure and dynamics is $\vec{\Pi}_{\vec{A}}$, calculated as,
\begin{align} \label{Pi}
\vec{\Pi}_\vec{A} = \frac{\partial\mathcal{L}}{\partial\dot{\vec{A}}} = -\bigg(\varepsilon_0 \vec{E} + \int_{0}^{\infty}d\omega \,& \alpha(\textbf{r}, \omega){\textbf{X}}_{\omega} \notag \\ 
&+ \frac{\alpha \theta}{\pi Z_0} \vec{B} + \vec{d}_{\rho}\bigg).
\end{align}
This can be re-written in terms of $\dot{\vec{A}}$, which is essential for the Legendre transformation to the Hamiltonian description in subsection~\ref{S2c},
\begin{align} \label{Adot}
&\dot{\vec{A}} = -\nabla \phi \notag \\
&+\frac{1}{\varepsilon_0} \left(\vec{\Pi}_\vec{A} + \int_{0}^{\infty}d\omega \, \alpha(\textbf{r}, \omega){\textbf{X}}_{\omega} + \frac{\alpha \theta}{\pi Z_0} \vec{B} + \vec{d}_{\rho}\right),
\end{align}
while the electric field can be written as, 
\begin{equation} \label{Efield}
{\vec{E}} = - \frac{1}{\varepsilon_0} \left(\vec{\Pi}_\vec{A} + \int_{0}^{\infty}d\omega \, \alpha(\textbf{r}, \omega){\textbf{X}}_{\omega} + \frac{\alpha \theta}{\pi Z_0} \vec{B} + \vec{d}_{\rho}\right).
\end{equation}
Given Eq.~\eqref{phiEOM}, then $\nabla \cdot \vec{\Pi}_\vec{A}=0$, meaning $\vec{\Pi}_\vec{A}$ is purely transverse and in classical electrodynamics would be given by $\vec{\Pi}_{\vec{A}}=\vec{\Pi}_{\vec{A}}^{\perp}=-\vec{D}^{\perp}$ for the case of no free charges, $\rho_{f}=0$. It is shown in Appendix~\ref{AppendixA} how exactly the $\vec{D}$ and $\vec{H}$ fields have been modified due to the non-reciprocal magneto-electric interaction. The important aspect here is the emergence of a hidden momentum contribution in Eq.~\eqref{Pi} that is proportional to $\vec{B}$, which would be often overlooked without such a canonical quantisation approach. Together, with the dipole moment density in Eq.~\eqref{Pi} an additional contribution to the interaction Hamiltonian will appear. Therefore, the next step is to transform the Lagrangian to a Hamiltonian description and quantise the system.

\subsection{Hamiltonian Quantisation} \label{S2c}

In order to rigorously quantise this system, it is imperative that the scalar potential, $\phi$ is eliminated as a dynamic variable as outlined in Ref.~\cite{philbin2010canonical} and described more generally in Ref.~\cite{weinberg1995quantum}. The constraint of $\nabla \cdot \vec{\Pi}_\vec{A}=0$ implies that for field quantisation Dirac brackets must be computed rather than Possion brackets. Therefore, along with some choice of gauge, $\phi$ can be eliminated, leaving $\vec{A}$ and $\vec{\Pi}_\vec{A}$ as the canonical variables of the electromagnetic field~\cite{weinberg1995quantum}. The Coulomb gauge is imposed, $\nabla \cdot \vec{A}=0$, such that $\phi$ is effectively eliminated with both $\vec{A}$ and $\vec{\Pi}_{\vec{A}}$ transverse, leading to the following canonical commutation relations~\cite{philbin2010canonical}
\begin{align}
\left[\hat{A}_i(\vec{r},t), \hat{\Pi}_{A,j}(\vec{r'},t) \right] &= i \hbar \delta_{ij}^{\perp}(\vec{r}-\vec{r'}), \label{APiCCR}\\
\left[\hat{A}_i(\vec{r},t), \hat{A}_{j}(\vec{r'},t) \right]&=0, \label{AACCR}\\
\left[\hat{\Pi}_{A,i}(\vec{r},t), \hat{\Pi}_{A,j}(\vec{r'},t) \right]&=0, \label{PiPiCCR}
\end{align}
where $\bm{\delta}^{\perp}(\vec{r} - \vec{r'})$ is the transverse delta function. Given that the fields $\hat{\vec{X}}_{\omega}$ and $\hat{\vec{\Pi}}_{\vec{X}_{\omega}}$ are unconstrained, they will obey the usual canonical commutation relations
\begin{align}\label{XPiXCCR}
\left[\hat{X}_{\omega,i}(\vec{r},t), \hat{\Pi}_{\vec{X}_{\omega'},j}(\vec{r'},t) \right] &= i \hbar \delta_{ij}(\vec{r}-\vec{r'}) \delta(\omega-\omega'),  \\
\left[\hat{X}_{\omega,i}(\vec{r},t), \hat{X}_{\omega',j}(\vec{r'},t) \right]&=0,   \\
\left[\hat{\Pi}_{\vec{X}_{\omega},i}(\vec{r},t), \hat{\Pi}_{\vec{X}_{\omega'},j}(\vec{r'},t) \right]&=0, 
\end{align}
with their magnetic reservoir oscillator counterparts, $\hat{\vec{Y}}_{\omega}$ and $\hat{\vec{\Pi}}_{\vec{Y}_{\omega}}$ obeying identical relations. Now the quantised Hamiltonian density can be written as, 
\begin{align}\label{H}
\hat{\mathcal{H}} &= \frac{1}{2}\left(\hat{\vec{\Pi}}_A \cdot \hat{\dot{\vec{A}}} + \hat{\dot{\vec{A}}} \cdot \hat{\vec{\Pi}}_A  \right) + \hat{\vec{\Pi}}_{\phi} \cdot \hat{\dot{\phi}} \notag \\ 
&+ \int_0^{\infty} d\omega\, \bigl(\hat{\vec{\Pi}}_{X_{\omega}} \cdot \hat{\dot{\vec{X}}}_{\omega} 
 + \hat{\vec{\Pi}}_{Y_{\omega}} \cdot \hat{\dot{\vec{Y}}}_{\omega} \bigl) - \hat{\mathcal{L}}, 
\end{align}
where $\hat{\vec{\Pi}}_{\phi}=0$ and the first two terms in brackets have been written in a symmetric form such that $\hat{\mathcal{H}}$ is Hermitian given that $\hat{\dot{\vec{A}}}$ features $\hat{\vec{B}}=\nabla \times \hat{\vec{A}}$~\eqref{Adot}, which will not commute with $\hat{\vec{\Pi}}_\vec{A}$. The Hamiltonian is cast into the form, $\hat{\mathcal{H}}=\hat{\mathcal{H}}_\mathrm{R}+\hat{\mathcal{H}}_\mathrm{NR}+\hat{\mathcal{H}}_\mathrm{I}$, where $\hat{\mathcal{H}}_\mathrm{R}$ is the usual reciprocal Hamiltonian structure derived in Ref.~\cite{philbin2010canonical}, $\hat{\mathcal{H}}_\mathrm{NR}$ corresponds to the field-only terms which feature the non-reciprocal coupling, $\theta(\vec{r})$, and $\hat{\mathcal{H}}_\mathrm{I}$ is the dipole-field interaction Hamiltonian density. 

The dynamical variables $\hat{\dot{\vec{A}}}$, $\hat{\dot{\vec{X}}}_{\omega}$, $\hat{\dot{\vec{Y}}}_{\omega}$ can now be written in terms of the canonical momenta via Eqs.~\eqref{Xpi}, \eqref{Ypi}, and \eqref{Adot}, then after some algebra the Hamiltonian, $\hat{H}=\int d^3 \vec{r}\, \hat{\mathcal{H}}$ can be partitioned as $\hat{{H}}=\hat{H}_\mathrm{R}+\hat{{H}}_\mathrm{NR}+\hat{{H}}_\mathrm{I}$, where:
\begin{align} 
& \hat{H}_R = \int d^3 \vec{r}\, \biggl(\frac{1}{2\varepsilon_0} \hat{\vec{\Pi}}_A^2 + \frac{1}{\varepsilon_0} \hat{\vec{\Pi}}_A \cdot \int_0^{\infty} d\omega \, \alpha(\vec{r},\omega) \hat{\vec{X}}_{\omega} \notag \\
& + \frac{1}{2}\int_0^{\infty} d\omega \, \bigl(\hat{\vec{\Pi}}_{X_{\omega}}^2 + \omega^2 \hat{\vec{X}}_{\omega}^2 + \hat{\vec{\Pi}}_{Y_{\omega}}^2 + \omega^2 \hat{\vec{Y}}_{\omega}^2  \bigl) \notag\\
& + \frac{1}{2\varepsilon_0}\int_0^{\infty} d\omega \, \alpha(\vec{r},\omega) \hat{\vec{X}}_{\omega} \cdot \int_0^{\infty} d\omega' \, \alpha(\vec{r},\omega') \hat{\vec{X}}_{\omega'}\notag \\
& + \frac{1}{2\mu_0} (\nabla \times \hat{\vec{A}})^2 - \int_0^{\infty} d\omega \, \beta(\vec{r},\omega) \hat{\vec{Y}}_{\omega} \cdot (\nabla \times \hat{\vec{A}}) \biggl), \label{HR}\\
& \hat{H}_\mathrm{NR} = -\int d^3 \vec{r}\, \biggl(\frac{\alpha \theta}{2\pi Z_0}(\hat{\vec{E}}\cdot\hat{\vec{B}} + \hat{\vec{B}} \cdot \hat{\vec{E}}) + \frac{\alpha^2 \theta^2}{2\mu_0 \pi^2}\hat{\vec{B}}^2 \biggl),\label{HNR}
\end{align}
and
\begin{equation}
\hat{H}_\mathrm{I} = -\int d^3 \vec{r}\, \biggl(\frac{1}{2\varepsilon_0}\hat{\vec{d}}_{\rho}^2 + \hat{\vec{d}}_{\rho} \cdot \hat{\vec{E}} + \frac{\alpha c \theta}{\pi}\hat{\vec{d}}_{\rho} \cdot \hat{\vec{B}} \biggl). \label{HI}
\end{equation}
There is a $\hat{\vec{\Pi}}_\vec{A} \cdot \nabla \hat{\phi}$ term that would feature in Eq.~\eqref{HR}, however this terms vanishes in the spatial integral given that $\hat{\vec{\Pi}}_\vec{A}$ is transverse and $\nabla \hat{\phi}$ is longitudinal. $\hat{H}_\mathrm{R}$ is just the total Hamiltonian that was derived in Ref.~\cite{philbin2010canonical} while $\hat{H}_\mathrm{NR}$ is the modification to the total Hamiltonian due to the non-reciprocal interaction mixing the electric and magnetic fields. The same diagonalisation procedure conveyed in \cite{philbin2010canonical} can be applied to $\hat{H}_0=\hat{H}_R + \hat{H}_\mathrm{NR}$ so that the entire system is diagonalised in terms of the polariton ladder operators, $\hat{\vec{f}}(\vec{r},\omega)$ and $\hat{\vec{f}}^{\dagger}(\vec{r},\omega)$. Owing to the $\delta$-function in the dipole density, $\hat{\vec{d}}_{\rho}$, the interaction Hamiltonian in Eq.~\eqref{HI} can be simplified to

\begin{align}\label{HI2}
& \hat{H}_\mathrm{I} = - \biggl(\frac{1}{2\varepsilon_0}\hat{\vec{d}}^2(\vec{r}_\mathrm{A},t) \delta(0) + \hat{\vec{d}}(\vec{r}_\mathrm{A},t) \cdot \hat{\vec{E}}(\vec{r}_\mathrm{A},t) \notag \\
&\quad\quad\quad\quad\quad+ \frac{\alpha c \theta(\vec{r}_\mathrm{A})}{\pi}\hat{\vec{d}}(\vec{r}_\mathrm{A},t) \cdot \hat{\vec{B}}(\vec{r}_\mathrm{A},t) \biggl).
\end{align}
The first term corresponds to a divergent dipole self-energy interaction which can be absorbed into the definition of the dipole moment, while the second term is the usual multipolar electric-dipole interaction. The last term is the novel interaction serving as the central result of our canonical quantisation procedure, which arises due to the magneto-electric coupling in the system. Normally, an electric dipole will not couple to the magnetic field. However due to the $\vec{E} \cdot \vec{B}$ coupling in Eq.~\eqref{L_NR}, we effectively see that the dipole couples to a mixture of the electric and magnetic fields, providing another route for interaction of the dipole that depends on the non-reciprocity. In particular, the electric dipole now couples also directly to the magnetic field, weighted by the magneto-electric coupling strength, $\theta(\vec{r}_\mathrm{A})$, at the position of the dipole. Importantly, this means that the additional interaction Hamiltonian is only non-vanishing when the dipole is within a medium with non-zero $\theta$. Therefore, a TSB-TI will not only affect the electromagnetic field, but also the coupling to other matter itself.

To reinforce the validity of this result, it is shown in Appendix~\ref{AppendixA} that the Lagrangian description used exactly reproduces the constitutive relations and inhomogeneous Helmholtz equation of the corresponding non-reciprocal system in \cite{crosse2015electromagnetic}. 

\section{Decay Rate in Non-Reciprocal Media} \label{S4}

As an example of how observables are modified by the interaction Hamiltonian in Eq.~\eqref{HI2}, we will calculate the spontaneous decay rate of an atom in the vicinity of a non-reciprocal medium. We will use Fermi's golden rule, which implicitly assumes the Born-Markov approximation for weak light-matter coupling with a bath that effectively relaxes back to equilibrium instantly,
\begin{equation} \label{Fermi}
\Gamma = \frac{2\pi}{\hbar} \sum_{f} \vert \langle f\vert \hat{H}_\mathrm{I} \vert i \rangle \vert^2 \delta(E_f - E_i),
\end{equation}
where $\vert i \rangle$, $\vert f \rangle $ are the initial and final states respectively during the decay process, and $E_i$, $E_f$ are the corresponding energies of those states. Having established the canonical form of the interaction Hamiltonian, the remaining calculation is most readily performed by first expressing the quantised electromagnetic field in terms of the light-matter polariton operators. This can be done in multiple ways. Given that the Hamiltonian $\hat{H}_0=\hat{H}_R + \hat{H}_\mathrm{NR}$ is quadratic in the canonical field operators, it is explicitly diagonalisable using the method outlined in Ref.~\cite{philbin2010canonical}, which is equivalent to the phenomenological approach outlined in Refs.~\cite{buhmann2012macroscopic, fuchs2017casimir}. In this, we first find the classical equation of motion for the electric field. As shown in Appendix~\ref{AppendixA}, this takes the form
\begin{align}\label{Helmholtz}
&\nabla \times \frac{1}{\mu(\vec{r},\omega)} \nabla \times \vec{E}(\vec{r},\omega) - \frac{\omega^2}{c^2}\varepsilon(\vec{r},\omega)\vec{E}(\vec{r},\omega) \notag \\
&\quad\quad\quad - i \frac{\alpha \omega}{\pi c} \nabla \theta(\vec{r}) \times \vec{E}(\vec{r},\omega)  = \mu_0 i \omega \vec{J}_f(\vec{r},\omega),
\end{align}
which is in agreement with Ref.~\cite{crosse2015electromagnetic} where
\begin{align}\label{Helmholtzcurrent}
\vec{J}_f(\vec{r},\omega) &= \vec{J}_d(\vec{r},\omega) + \vec{J}_\mathrm{N}(\vec{r},\omega), 
\end{align}
where $\vec{J}_d$ is the dipole current and $\vec{J}_N$ is the noise current associated with the optical medium. We can then find the desired expression for the quantised electric field $\hat{\vec{E}}(\vec{r},\omega)$ in terms of polaritons by promoting the noise current $\vec{J}_N(\vec{r},\omega)$ (equivalently the polarisation/magnetisation) to operators, where
\begin{align} \label{JN}
\hat{\vec{J}}_N(\vec{r}, \omega) = \sqrt{\frac{\hbar \omega}{\pi}}\int d^3 \vec{r'}\, \vec{R}(\vec{r}, \vec{r'}, \omega) \cdot \hat{\vec{f}}(\vec{r'}, \omega),
\end{align}
where the tensor, $\vec{R}$ and the corresponding quantisation procedure are described further in Appendix~\ref{AppendixB}. As a consequence, we have the quantised electric field
\begin{align} \label{EfreqQ}
&\hat{\vec{E}}(\vec{r}, \omega) = \mu_0 i \omega \sqrt{\frac{\hbar \omega}{\pi}} \notag \\
& \times \int d^3 \vec{r'} \int d^3 \vec{s}\, \vec{G}(\vec{r}, \vec{r'}, \omega) \cdot \vec{R}(\vec{r'}, \vec{s}, \omega) \cdot \hat{\vec{f}}(\vec{s}, \omega),
\end{align}
with the magnetic field following from Maxwell's equations
\begin{align} \label{BfreqQ}
&\hat{\vec{B}}(\vec{r}, \omega) = \mu_0 \sqrt{\frac{\hbar \omega}{\pi}} \notag \\
& \times \int d^3 \vec{r'} \int d^3 \vec{s}\, [\nabla_{\vec{r}} \times \vec{G}(\vec{r}, \vec{r'}, \omega)] \cdot \vec{R}(\vec{r'}, \vec{s}, \omega) \cdot \hat{\vec{f}}(\vec{s}, \omega).
\end{align}
The polariton annihilation and creation operators, $\hat{\vec{f}}(\vec{r},\omega)$ and $\hat{\vec{f}}^{\dagger}(\vec{r},\omega)$ obey the usual bosonic canonical commutation relations~\cite{buhmann2012macroscopic}
\begin{align} \label{fCCR}
\left[\hat{f}_i(\vec{r},\omega), \hat{f}^{\dagger}_j(\vec{r'},\omega') \right] = \delta_{ij} \delta(\vec{r}-\vec{r'})\delta(\omega - \omega'),
\end{align}
while the dynamics of the polaritons is given by the Hamiltonian
\begin{align} \label{HMQED}
\hat{H}_0=\int d^3 \vec{r} \int_0^{\infty} d\omega \, \hbar \omega\, \hat{\vec{f}}^{\dagger}(\vec{r},\omega) \cdot \hat{\vec{f}}(\vec{r},\omega).
\end{align}
Excitations of this polariton field can be expressed in terms of the Fock basis for a vector-valued excitation given by, $\hat{\vec{f}}^{\dagger}(\vec{r},\omega) |\vec{0} \rangle = |\vec{1}(\vec{r},\omega) \rangle$. For simplicity, a two-level atom in the electric-dipole approximation will be assumed to initially be in its excited state, $|e\rangle$, with the medium-field bath initially in the vacuum state, $|\vec{0}\rangle$, where the energy spacing between the atomic ground and excited states is $\hbar \omega_\mathrm{A}$. Therefore, Eq.~\eqref{Fermi} can be expanded as
\begin{align} \label{Fermi expanded}
 \Gamma = & \frac{2\pi}{\hbar}\int d^3\vec{r''} \int_{0}^{\infty} d\omega \, \vert \langle \vec{1}(\vec{r''},\omega)\vert \langle g \vert \, \biggl(\hat{\vec{d}}(\vec{r}_\mathrm{A},t) \cdot \hat{\vec{E}}(\vec{r}_\mathrm{A},t) \notag \\
& + \frac{\alpha c \theta(\vec{r}_\mathrm{A})}{\pi}\hat{\vec{d}}(\vec{r}_\mathrm{A},t) \cdot \hat{\vec{B}}(\vec{r}_\mathrm{A},t)\, \biggl) \vert e \rangle \vert \vec{0} \rangle \vert^2 \delta(\hbar \omega - \hbar \omega_\mathrm{A}).
\end{align}
We can partition this as
\begin{align}\label{Fermi expanded 2}
 &\Gamma = \Gamma_\mathrm{E} + \Gamma_\mathrm{EB} + \Gamma_\mathrm{B} \notag \\ 
 &= \frac{2\pi}{\hbar}\int d^3\vec{r''} \int_{0}^{\infty} d\omega \, \bigl(M_\mathrm{E} + M_\mathrm{EB} + M_\mathrm{B} \bigl) \delta(\hbar \omega - \hbar \omega_\mathrm{A}),
\end{align}
with
\begin{align}
& M_\mathrm{E} = \vert \langle \vec{1}(\vec{r}'',\omega)\vert \langle g \vert \, \hat{\vec{d}} \cdot \hat{\vec{E}} \, \vert e \rangle \vert \vec{0} \rangle \vert^2, \label{gammaE}\\
& M_\mathrm{EB} = \frac{2 \alpha c \theta}{\pi} \notag\\
& \times \mathrm{Re}\bigl(\langle \vec{1}(\vec{r}'',\omega)\vert \langle g \vert \, \hat{\vec{d}} \cdot \hat{\vec{E}} \, \vert e \rangle \vert \vec{0} \rangle \langle \vec{0}\vert \langle e \vert \, \hat{\vec{d}} \cdot \hat{\vec{B}}\, \vert g \rangle \vert \vec{1}(\vec{r}'',\omega) \rangle \bigl), \label{gammaEB} \\
& M_\mathrm{B} = \frac{\alpha^2 c^2 \theta^2}{\pi^2} \vert \langle \vec{1}(\vec{r}'',\omega)\vert \langle g \vert \, \hat{\vec{d}} \cdot \hat{\vec{B}}\, \vert e \rangle \vert \vec{0} \rangle \vert^2,\label{gammaB}
\end{align}
such that only two matrix elements, \mbox{$\langle \vec{1}(\vec{r}'',\omega)\vert \langle g \vert \, \hat{\vec{d}} \cdot \hat{\vec{E}} \, \vert e \rangle \vert \vec{0} \rangle$} and \mbox{$\langle \vec{1}(\vec{r}'',\omega)\vert \langle g \vert \, \hat{\vec{d}} \cdot \hat{\vec{B}}\, \vert e \rangle \vert \vec{0} \rangle $} need to be explicitly calculated in order to determine the full decay rate. Equation~\eqref{gammaE} corresponds to the decay rate of the usual electric-dipole coupling, while the terms proportional to $\theta(\vec{r}_\mathrm{A})$ corresponds to the cross-interaction decay rate between $\hat{\vec{E}}$ and $\hat{\vec{B}}$, and the last term proportional to $\theta^2(\vec{r}_\mathrm{A})$ corresponds to the purely magnetic interaction decay rate.

The non-vanishing dipole matrix elements are given by $\langle e \vert \hat{\vec{d}}(\vec{r}_\mathrm{A},t) \vert g \rangle = \vec{d}^{\uparrow}(\vec{r}_\mathrm{A}) e^{i \omega_\mathrm{A} t}$ and its complex conjugate $\langle g \vert \hat{\vec{d}}(\vec{r}_\mathrm{A},t) \vert e \rangle = \vec{d}^{\downarrow}(\vec{r}_\mathrm{A}) e^{-i \omega_\mathrm{A} t}$. Following Eqs.~\eqref{EfreqQ} and~\eqref{BfreqQ} the time-dependent electric and magnetic fields can be written as
\begin{align}\label{E}
& \hat{\vec{E}}(\vec{r}_\mathrm{A},t) = \int d^3\vec{r'}\int d^3\vec{s} \int_{0}^{\infty} d\omega' \, \mu_0 i \omega' \sqrt{\frac{\hbar \omega'}{\pi}} \notag \\ 
& \times \biggl( \vec{G}(\vec{r}_\mathrm{A},\vec{r'},\omega')
 \cdot \vec{R}(\vec{r'},\vec{s},\omega') \cdot \hat{\vec{f}}(\vec{s},\omega')e^{-i\omega't} \notag \\
& - \vec{G}^{*}(\vec{r}_\mathrm{A},\vec{r'},\omega') \cdot \vec{R}^{*}(\vec{r'},\vec{s},\omega') \cdot \hat{\vec{f}}^{\dagger}(\vec{s},\omega')e^{i\omega't} \biggl),
\end{align}
and
\begin{align} \label{B}
& \hat{\vec{B}}(\vec{r}_\mathrm{A},t) = \int d^3\vec{r'}\int d^3\vec{s} \int_{0}^{\infty} d\omega' \, \mu_0 \sqrt{\frac{\hbar \omega'}{\pi}} \notag \\
& \times \biggl(\left[\overrightarrow{\nabla}_{\vec{r}_\mathrm{A}} \times \vec{G}(\vec{r}_\mathrm{A},\vec{r'},\omega')\right]
 \cdot \vec{R}(\vec{r'},\vec{s},\omega') \cdot \hat{\vec{f}}(\vec{s},\omega')e^{-i\omega't} \notag \\
& + \left[\overrightarrow{\nabla}_{\vec{r}_\mathrm{A}} \times \vec{G}^{*}(\vec{r}_\mathrm{A},\vec{r'},\omega')\right] \cdot \vec{R}^{*}(\vec{r'},\vec{s},\omega') \cdot \hat{\vec{f}}^{\dagger}(\vec{s},\omega')e^{i\omega't} \biggl).
\end{align}
The matrix, $\vec{G}$ is the electromagnetic Green's tensor for the inhomogeneous Helmholtz equation (\ref{AppendixA}) while the matrix, $\vec{R}$ is defined as 
\begin{equation} \label{R}
\int d^3\vec{r''}\, \vec{R}(\vec{s},\vec{r''},\omega) \cdot \vec{R}^{\dagger}(\vec{r'},\vec{r''},\omega) = \mathcal{R}e\vec{Q}(\vec{s},\vec{r'},\omega),
\end{equation}
where $\mathcal{R}e\vec{Q}$ is the generalised real part (Hermitian part), 
\begin{equation} \label{GenRe}
\mathcal{R}e\vec{Q}(\vec{r},\vec{r'},\omega) = \frac{1}{2}\left(\vec{Q}(\vec{r},\vec{r'},\omega) + \vec{Q}^{\dagger}(\vec{r'},\vec{r},\omega) \right),
\end{equation}
of the conductivity tensor, $\vec{Q}$ (\ref{AppendixB}) from Ref.~\cite{buhmann2012macroscopic}. The relevant matrix elements in Eqs.~\eqref{gammaE}-\eqref{gammaB} can be calculated as follows:
\begin{align}\label{dE}
& \langle \vec{1}(\vec{r''},\omega)\vert \langle g \vert \, \hat{\vec{d}}(\vec{r}_\mathrm{A},t) \cdot \hat{\vec{E}}(\vec{r}_\mathrm{A},t) \, \vert e \rangle \vert \vec{0} \rangle \notag \\
& = \langle g \vert \hat{\vec{d}}(\vec{r}_\mathrm{A},t) \vert e \rangle \cdot \langle \vec{1}(\vec{r''},\omega)\vert \hat{\vec{E}}(\vec{r}_\mathrm{A},t) \vert \vec{0} \rangle  \notag\\
& = e^{-i \omega_\mathrm{A} t}\vec{d}^{\downarrow} \cdot \int d^3\vec{r'}\int d^3\vec{s} \int_{0}^{\infty} d\omega' \, \mu_0 i \omega' \sqrt{\frac{\hbar \omega'}{\pi}}  \notag\\ 
& \times \langle \vec{0} \vert \hat{\vec{f}}(\vec{r''},\omega) \cdot \biggl( \vec{G}(\vec{r}_\mathrm{A},\vec{r'},\omega')
 \cdot \vec{R}(\vec{r'},\vec{s},\omega') \cdot \hat{\vec{f}}(\vec{s},\omega')e^{-i\omega't} \notag \\
& - \vec{G}^{*}(\vec{r}_\mathrm{A},\vec{r'},\omega') \cdot \vec{R}^{*}(\vec{r'},\vec{s},\omega') \cdot \hat{\vec{f}}^{\dag}(\vec{s},\omega')e^{i\omega't} \biggl) \vert \vec{0} \rangle,
\end{align}
where the first term in Eq.~\eqref{dE} will vanish as $\hat{\vec{f}}(\vec{s},\omega')\vert \vec{0} \rangle = 0$ while the second term can be re-written using Eq.~\eqref{fCCR} as
\begin{align}\label{dE2}
& \langle \vec{1}(\vec{r''},\omega)\vert \langle g \vert \, \hat{\vec{d}}(\vec{r}_\mathrm{A},t) \cdot \hat{\vec{E}}(\vec{r}_\mathrm{A},t) \, \vert e \rangle \vert \vec{0} \rangle \notag \\
& = -e^{-i \omega_\mathrm{A} t} d_l^{\downarrow} \int d^3\vec{r'}\int d^3\vec{s} \int_{0}^{\infty} d\omega' \, \mu_0 i \omega' \sqrt{\frac{\hbar \omega'}{\pi}}\notag\\ 
& \times G_{lk}^{*}(\vec{r}_\mathrm{A},\vec{r'},\omega') R_{kj}^{*}(\vec{r'},\vec{s},\omega')e^{i\omega't} \notag\\
& \times \langle \vec{0} \vert \hat{f}_j^{\dagger}(\vec{s},\omega')\hat{f}_i(\vec{r''},\omega) + \delta_{ij}\delta(\omega - \omega')\delta(\vec{r''} - \vec{s}) \vert \vec{0} \rangle,
\end{align}
which simplifies to
\begin{align} \label{dE3}
& \langle \vec{1}(\vec{r''},\omega)\vert \langle g \vert \, \hat{\vec{d}}(\vec{r}_\mathrm{A},t) \cdot \hat{\vec{E}}(\vec{r}_\mathrm{A},t) \, \vert e \rangle \vert \vec{0} \rangle = -\mu_0 i \omega \sqrt{\frac{\hbar \omega}{\pi}} \notag  \\
& \!\times \vec{d}^{\downarrow} \cdot \! \int d^3\vec{r'} \, \vec{G}^{*}(\vec{r}_\mathrm{A},\vec{r'},\omega) \cdot \vec{R}^{*}(\vec{r'},\vec{r''},\omega)e^{i(\omega-\omega_\mathrm{A}) t}.
\end{align}
The same procedure outlined for Eqs.~\eqref{dE}--\eqref{dE3} can be applied for the matrix element $\langle \vec{1}(\vec{r''},\omega)\vert \langle g \vert \, \hat{\vec{d}} \cdot \hat{\vec{B}} \, \vert e \rangle \vert \vec{0} \rangle$, giving:
\begin{align}\label{dB}
& \langle \vec{1}(\vec{r''},\omega)\vert \langle g \vert \, \hat{\vec{d}}(\vec{r}_\mathrm{A}) \cdot \hat{\vec{B}}(\vec{r}_\mathrm{A},t) \, \vert e \rangle \vert \vec{0} \rangle = \mu_0 \sqrt{\frac{\hbar \omega}{\pi}} \vec{d}^{\downarrow} \,\times \notag \\
& \int d^3\vec{r'} \, \left[\overrightarrow{\nabla}_{\vec{r}_\mathrm{A}} \times \vec{G}^{*}(\vec{r}_\mathrm{A},\vec{r'},\omega)\right] \cdot \vec{R}^{*}(\vec{r'},\vec{r''},\omega)e^{i(\omega-\omega_\mathrm{A}) t}.
\end{align}
Equation~\eqref{dE3} can now be used, along with its complex conjugate to derive the purely electric decay rate, $\Gamma_\mathrm{E}$.
\begin{align}\label{GammaE}
 \Gamma_\mathrm{E} = & \frac{2\pi}{\hbar}\int d^3\vec{r}'' \int d^3\vec{r}^{'} \int d^3\vec{s}\int_{0}^{\infty} d\omega \, (-\mu_0 i \omega)(\mu_0 i \omega) \frac{\hbar \omega}{\pi} \notag \\
 & \times \left(\vec{d}^{\uparrow} \cdot \vec{G}(\vec{r}_\mathrm{A},\vec{s},\omega) \cdot \vec{R}(\vec{s},\vec{r}'',\omega) \right)\notag  \\
 & \times \left(\vec{d}^{\downarrow} \cdot \vec{G}^{*}(\vec{r}_\mathrm{A},\vec{r}^{'},\omega) \cdot \vec{R}^{*}(\vec{r}^{'},\vec{r}'',\omega)\right) \delta(\hbar \omega - \hbar \omega_\mathrm{A}).
\end{align}
The integrand of Eq.~\eqref{GammaE} is first re-written using $(\vec{a} \cdot \vec{B}) \cdot (\vec{c} \cdot \vec{D}) = \vec{a} \cdot (\vec{B} \cdot \vec{D}^T) \cdot \vec{c}$, 
\begin{align}\label{GammaE2}
 & \Gamma_\mathrm{E} = 2\mu_0^2 \vec{d}^{\uparrow} \cdot \int d^3\vec{r''} \int d^3\vec{r'} \int d^3\vec{s} \notag \\
 & \times\int_{0}^{\infty} d\omega \, \omega^3 \delta(\hbar \omega - \hbar \omega_\mathrm{A}) \left( \vec{G}(\vec{r}_\mathrm{A},\vec{s},\omega) \cdot \vec{R}(\vec{s},\vec{r''},\omega) \right. \notag\\ 
 & \left. \cdot \vec{R}^{\dagger}(\vec{r'},\vec{r''},\omega) \cdot \vec{G}^{\dagger}(\vec{r}_\mathrm{A},\vec{r'},\omega) \right) \cdot \vec{d}^{\downarrow},
\end{align}
which can be drastically simplified by first applying Eq.~\eqref{R}, then the following relation between $\vec{G}$ and $\vec{Q}$ from Ref.~\cite{buhmann2012macroscopic} can be applied
\begin{align}\label{Q}
& \mu_0 \omega \int d^3\vec{s} \int d^3\vec{r'}\, \bigl(\vec{G}(\vec{r}_\mathrm{A},\vec{s},\omega) \cdot \mathcal{R}e\vec{Q}(\vec{s},\vec{r'},\omega)  \notag\\
&\cdot \vec{G}^{\dagger}(\vec{r}_\mathrm{A},\vec{r'},\omega) \bigl) = \mathcal{I}m\vec{G}(\vec{r}_\mathrm{A},\vec{r}_\mathrm{A},\omega),
\end{align}
where $\mathcal{I}m\vec{G}$ is the generalised imaginary part (anti-Hermitian part),
\begin{equation} \label{GenIm}
\mathcal{I}m\vec{G}(\vec{r},\vec{r'},\omega) = \frac{1}{2i}\left(\vec{G}(\vec{r},\vec{r'},\omega) - \vec{G}^{\dagger}(\vec{r'},\vec{r},\omega) \right),
\end{equation}
of the Green's tensor. Then, the purely electric decay rate contribution is calculated as
\begin{equation} \label{GammaE3}
\Gamma_\mathrm{E} = \frac{2\mu_0 \omega_\mathrm{A}^2}{\hbar}\vec{d}^{\uparrow} \cdot \mathcal{I}m\vec{G}(\vec{r}_\mathrm{A},\vec{r}_\mathrm{A},\omega_\mathrm{A}) \cdot \vec{d}^{\downarrow},
\end{equation}
which to first-order in the atomic frequency shift is in agreement with the decay rate derived in Ref.~\cite{fuchs2017casimir} for non-reciprocal media. Furthermore, for the case of a real-valued dipole moment, Eq.~\eqref{GammaE3} simplifies to
\begin{align} \label{GammaE4}
\Gamma_\mathrm{E} = \frac{2\mu_0 \omega_\mathrm{A}^2}{\hbar}\vec{d} \cdot \mathrm{Im}\vec{G}(\vec{r}_\mathrm{A},\vec{r}_\mathrm{A},\omega_\mathrm{A}) \cdot \vec{d},
\end{align}
which now features the standard imaginary part, Im, rather than the generalised imaginary part, $\mathcal{I}m$ seen in Eq.~\eqref{GammaE3}. The structure of Eq.~\eqref{GammaE4} is now identical to the decay rate for reciprocal media, however, as will be discussed further in section~\ref{S5}, the Green's tensor, $\vec{G}$ will be modified due to the magneto-electric coupling, $\theta$. As a consequence, there will be variations to the decay rate compared to the reciprocal case~\cite{fuchs2017casimir} and $\vec{G}$ may feature off-diagonal components that can also contribute to the decay rate.

The purely magnetic decay rate, $\Gamma_\mathrm{B}$ can be calculated using Eq.~\eqref{dB}
\begin{align}\label{GammaB}
& \Gamma_\mathrm{B} = \frac{2\pi \mu_0^2 \alpha^2 c^2 \theta^2}{\pi^2 \hbar}\int d^3\vec{r''} \int d^3\vec{r'} \int d^3\vec{s}\int_{0}^{\infty} d\omega \, \frac{\hbar \omega}{\pi} \notag \\
 & \times \left(\vec{d}^{\uparrow} \cdot \overrightarrow{\nabla}_{\vec{r}_\mathrm{A}} \times \vec{G}(\vec{r}_\mathrm{A},\vec{s},\omega) \cdot \vec{R}(\vec{s},\vec{r''},\omega) \right) \notag \\
 & \times \left(\vec{d}^{\downarrow} \cdot \overrightarrow{\nabla}_{\vec{r}_\mathrm{A}} \times \vec{G}^{*}(\vec{r}_\mathrm{A},\vec{r'},\omega) \cdot \vec{R}^{*}(\vec{r'},\vec{r''},\omega)\right) \delta(\hbar \omega - \hbar \omega_\mathrm{A}),
\end{align}
and by following the same line of reasoning from Eqs.~\eqref{GammaE}-\eqref{GammaE3}, and making use of $\left(\overrightarrow{\nabla}_{\vec{r}} \times \vec{G}(\vec{r},\vec{r'},\omega) \right)^{\dagger} =  \vec{G}^{\dagger}(\vec{r},\vec{r'},\omega) \times \overleftarrow{\nabla}_{\vec{r}}$, $\Gamma_\mathrm{B}$ can be written as
\begin{align}\label{GammaB2}
&\Gamma_\mathrm{B} =  \frac{2\mu_0 \alpha^2 c^2}{\pi^2 \hbar} \theta^2(\vec{r}_\mathrm{A}) \notag  \\
& \times \left[ \lim_{\vec{r}\to \vec{r}_\mathrm{A}} \lim_{\vec{r'}\to \vec{r}_\mathrm{A}}\vec{d}^{\uparrow}  \cdot \overrightarrow{\nabla}_{\vec{r}} \times \, \mathcal{I}m\vec{G}(\vec{r},\vec{r'},\omega_\mathrm{A}) \, \times \overleftarrow{\nabla}_{\vec{r'}} \cdot \,\vec{d}^{\downarrow} \right].
\end{align}
Equation~\eqref{GammaB2} has the same structure as the decay rate due to magnetic dipole coupling, $\hat{\vec{m}}\cdot \hat{\vec{B}}$~\cite{buhmann2013dispersion}, except we have dipole coupling to the magnetic field, weighted by ${\alpha c \theta}/{\pi}$. Lastly, by utilising Eqs.~\eqref{dE}-\eqref{dB}, the cross decay rate, $\Gamma_\mathrm{EB}$ can be written as
\begin{align}\label{GammaEB}
& \Gamma_\mathrm{EB} =  \frac{2\pi}{\hbar}\frac{ 2\alpha c \theta}{\pi} \notag \\
& \times \mathrm{Re} \biggl( \int d^3\vec{r''}\int d^3\vec{r'}\int d^3\vec{s}\int_{0}^{\infty} d\omega \, \mu_0(-\mu_0 i \omega)\frac{\hbar \omega}{\pi} \notag \\
& \times \left(\vec{d}^{\uparrow} \cdot \overrightarrow{\nabla}_{\vec{r}_\mathrm{A}} \times \vec{G}(\vec{r}_\mathrm{A},\vec{s},\omega) \cdot \vec{R}(\vec{s},\vec{r''},\omega)\right)\notag  \\
& \times \left(\vec{d}^{\downarrow} \cdot \vec{G}^{*}(\vec{r}_\mathrm{A},\vec{r'},\omega) \cdot \vec{R}^{*}(\vec{r'},\vec{r''},\omega)\right) \delta(\hbar \omega - \hbar \omega_\mathrm{A}) \biggl).
\end{align}
The same reasoning presented in the derivations for $\Gamma_\mathrm{E}$ and $\Gamma_\mathrm{B}$ can be applied to simplify $\Gamma_\mathrm{EB}$ as
\begin{align}\label{GammaEB3}
& \Gamma_\mathrm{EB} =  \frac{4 \alpha c \mu_0 \omega_\mathrm{A} \theta(\vec{r}_\mathrm{A})}{ \pi \hbar} \notag \\
& \times \lim_{\vec{r}\to \vec{r}_\mathrm{A}} \mathrm{Im} \left[\vec{d}^{\uparrow} \cdot \bigl(\overrightarrow{\nabla}_{\vec{r}} \times \mathcal{I}m \vec{G}(\vec{r},\vec{r}_\mathrm{A},\omega_\mathrm{A}) \bigl) \cdot \, \vec{d}^{\downarrow} \right].
\end{align}
which, we note, depends on the standard imaginary part of the curl of the generalised imaginary part of $\vec{G}$. A similar structure to Eq.~\eqref{GammaEB3} can be seen in Ref.~\cite{rapp2025purcell} for the chiral cross decay rate in reciprocal media, which features the standard imaginary part throughout. The cross decay rate contribution there is only non-vanishing due to the consideration of complex-valued electric and magnetic dipole moments required for chiral systems. The key difference here is that even for real-valued dipole moments we have 
\begin{align}\label{GammaEBReal}
& \Gamma_\mathrm{EB} =  \frac{4 \alpha c \mu_0 \omega_\mathrm{A} \theta(\vec{r}_\mathrm{A})}{ \pi \hbar} \notag \\
& \times \lim_{\vec{r}\to \vec{r}_\mathrm{A}} \vec{d}\cdot \mathrm{Im} \left[\overrightarrow{\nabla}_{\vec{r}} \times \mathcal{I}m \vec{G}(\vec{r},\vec{r}_\mathrm{A},\omega_\mathrm{A}) \right] \cdot \vec{d},
\end{align}
where $\Gamma_\mathrm{EB}$ is non-vanishing due to the off-diagonal structure of the Green's tensor. The diagonal contributions of $\Gamma_{EB}$ in Eq.~\eqref{GammaEBReal} will vanish as the generalised imaginary part will behave as the usual imaginary part for those terms. Therefore, the only terms contributing to $\Gamma_\mathrm{EB}$ are the off-diagonal components of $\vec{G}$, providing a unique contribution to the total decay rate that will be investigated further in section~\ref{S5}.

\section{Numerical Calculation of Spontaneous Decay Rate} \label{S5}

\subsection{Planar Cavity Green's Tensor}

\begin{figure}[htb]
	\centering
	\includegraphics[width=9cm, height=5.5cm]{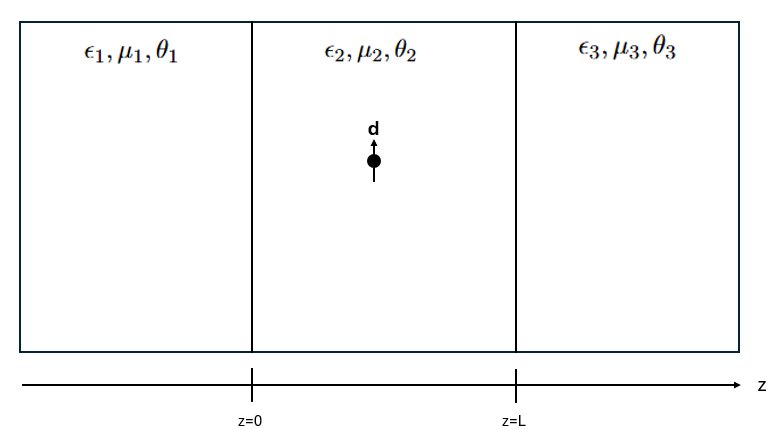}
	\caption{\label{Cavity} Planar cavity setup for emitter in central region with dipole moment, $\vec{d}=(d_x, d_y, d_z).$
	}
\end{figure}

To explicitly demonstrate the effect of the novel decay rate terms in Eqs. \eqref{GammaB2} and \eqref{GammaEB3}, an example calculation will be presented for the TSB-TI parallel-plate cavity. As shown in Fig.~\ref{Cavity}
we consider a three-layer system with permittivities, $\varepsilon_i$, permeabilities, $\mu_i$, and magneto-electric couplings, $\theta_i$ for each region, $i$, with a single emitter having dipole moment, $\vec{d}$ in the central region. The general structure for the Green's tensor is~\cite{crosse2015electromagnetic}:
\begin{align}\label{NRGreen}
&\vec{G}(\vec{r},\vec{r'},\omega) = \frac{i}{8\pi^2} \notag \\
&\times \int d^2\vec{k}_p\frac{\mu(\vec{r'}) \vec{C(\vec{r}, \vec{r'}) : \vec{F}(z,z')} }{k_z k_p^2}e^{i \vec{k}_p \cdot (\vec{r_p}-\vec{r_p'})},
\end{align}
where the $:$ operation indicates the Frobenius inner product. Here, $\vec{C}$ is a $2\times 2$ matrix of dyadic operators, which correspond to the polarisation vectors typically used in Green's tensor calculations
\begin{equation}
\vec{C}(\vec{r}, \vec{r'})=
\begin{bmatrix}
\vec{m}^*(\vec{r}) \otimes \vec{m}(\vec{r'}) & \vec{n}^*(\vec{r}) \otimes \vec{m}(\vec{r'}) \\
\vec{m}^*(\vec{r}) \otimes \vec{n}(\vec{r'}) & \vec{n}^*(\vec{r}) \otimes \vec{n}(\vec{r'}) 
\end{bmatrix},
\end{equation}
with,
\begin{align}
& \vec{m}(\vec{r}) = i \nabla_{\vec{r}} \times \hat{z}\,, \label{mvec}\\
& \vec{n}(\vec{r})=\frac{1}{k} \nabla_{\vec{r}} \times \nabla_{\vec{r}} \times \hat{z}\,. \label{nvec}
\end{align}
Finally, $\vec{F}$ represents the response to an initial source electric field in the central layer and is given by
\begin{align} \label{Ftensor}
& \vec{F}(z,z') = e^{i k_z z} \vec{M}_{-+} \cdot \left[e^{-i k_z z'} \vec{I} + e^{i k_z z'} \vec{R}_{-}\right]  \notag \\
&+e^{-i k_z z} \vec{M}_{+-} \cdot \left[e^{-i k_z (z'-2L)}\vec{R}_{+} + e^{i k_z (z'+2L)}\vec{R}_{+} \cdot \vec{R}_{-}  \right],
\end{align}
with,
\begin{align}
& \vec{M}_{ij}=\left[\vec{I}-e^{2ik_z L}\, \vec{R}_i \cdot \vec{R}_j \right]^{-1}, \label{Multreflect} \\
& \vec{R}_{i} = \begin{bmatrix}
R_\mathrm{TE, TE} & R_\mathrm{TE, TM} \\
R_\mathrm{TM, TE} & R_\mathrm{TM, TM}
\end{bmatrix}, \label{ReflectMat}
\end{align}
where $R_\mathrm{TE,TE}$ and $R_\mathrm{TM,TM}$ are the usual Fresnel reflection coefficients for the interfaces, while the $R_\mathrm{TE,TM}$ and $R_\mathrm{TM,TE}$ coefficients correspond to reflections in which the polarisation can transform between TE and TM polarisations, purely due to the inclusion of a non-reciprocal interaction, modifying the constitutive relations. The $-$ and $+$ indices in Eq.~\eqref{Ftensor} represent the coefficient matrix upon reflections from the left or right wall in Fig.~\ref{Cavity} respectively. For example, the reflection coefficients for reflections at the left wall are given by,
\begin{align}
& R_\mathrm{TE, TE} = \frac{(\mu_2 k_{z,1} - \mu_1 k_{z,2})\Omega_{\varepsilon} - k_{z,1}k_{z,2}\Delta^2}{(\mu_2 k_{z,1} + \mu_1 k_{z,2})\Omega_{\varepsilon} + k_{z,1}k_{z,2}\Delta^2}, \label{TETE}\\
& R_\mathrm{TE, TM}=\frac{-2\mu_2 n_1 k_{z,1}k_{z,2} \Delta}{(\varepsilon_2 k_{z,1} + \varepsilon_1 k_{z,2})\Omega_{\mu} + k_{z,1}k_{z,2}\Delta^2}, \label{TETM}\\
& R_\mathrm{TM, TE} = \frac{-2\mu_2 n_1 k_{z,1}k_{z,2} \Delta}{(\mu_2 k_{z,1} + \mu_1 k_{z,2})\Omega_{\varepsilon} + k_{z,1}k_{z,2}\Delta^2}, \label{TMTE}\\
& R_\mathrm{TM, TM}=\frac{(\varepsilon_2 k_{z,1} - \varepsilon_1 k_{z,2})\Omega_{\mu} - k_{z,1}k_{z,2}\Delta^2}{(\varepsilon_2 k_{z,1} + \varepsilon_1 k_{z,2})\Omega_{\mu} + k_{z,1}k_{z,2}\Delta^2}, \label{TMTM}
\end{align}
where,
\begin{align}
& \Delta = \frac{\alpha \mu_1 \mu_2}{\pi}(\theta_2 - \theta_1), \label{Delta} \\
& \Omega_{\mu} = \mu_1 \mu_2 (\mu_2 k_{z,1} + \mu_1 k_{z,2}), \label{OmegaMu}\\
& \Omega_{\varepsilon} = \mu_1 \mu_2 (\varepsilon_2 k_{z,1} + \varepsilon_1 k_{z,2}), \label{OmegaEp}
\end{align}
and $k_{z,i}$ represents the z-component of the wavevector in region i, while $\vec{k_p}$ is the in-plane wavevector, and $k_i$ is the absolute magnitude of the wavevector in region i
\begin{align}
& \vec{k_p}=k_x \hat{\vec{x}} + k_y \hat{\vec{y}}, \label{kp}\\
& k_{z,i} = \sqrt{k_i^2 - k_p^2}, \label{kz}\\
&k_i=\frac{n_i \omega}{c}. \label{ktotal}
\end{align}
The non-reciprocity based alteration to the constitutive equations essentially marks the dimensionless non-reciprocity parameter, $\theta$, as another medium property like $\varepsilon$ and $\mu$. This allows for the mixing of electric and magnetic fields, meaning that the TE and TM polarisations are no longer preserved upon reflection and transmission, hence the additional reflection coefficients and required matrix structure.

\subsection{Numerical Results}

With the Green's tensor laid out as before, the decay rate can be calculated for various parameters. As a simple example to demonstrate the effect the non-reciprocal terms have, it will be assumed that $\varepsilon=\mu=1$ throughout each layer, and that $\theta_1=\theta_3$. This will highlight any variations in the decay rate purely due to the non-reciprocity and will mean that the corresponding reciprocal decay rate is just the vacuum decay rate, $\Gamma_0={\mu_0 |\vec{d}|^2 \omega_\mathrm{A}^3}/(3\pi \hbar c)$. Additionally, for the calculations it is assumed that $\vec{d}=(d,0,0)$ with $L=10\mu m$ for the cavity length.

\begin{figure}[htb]
	\centering
	\includegraphics[width=8.5cm, height=5.5cm]{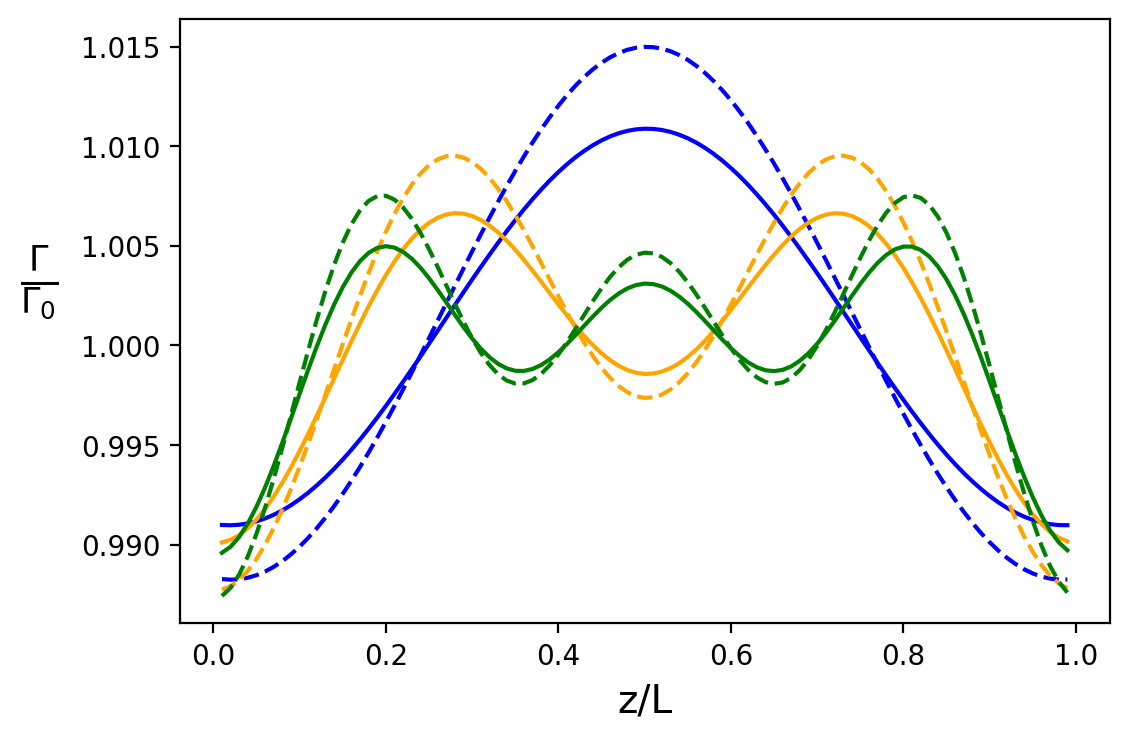}
	\caption{\label{GammaE vs GammaT (th2_10)} Position dependent decay rate normalised by $\Gamma_0$ ($\theta_1=100,\, \theta_2=10$). $\Gamma_\mathrm{E}$ (solid), $\Gamma_T$ (dashed).
    $\frac{\omega_\mathrm{A} L}{\pi c} = (1,2,3)$ for (blue, orange, green) lines respectively.
	}
\end{figure}

Fig.~\ref{GammaE vs GammaT (th2_10)} highlights the variation in decay rate between the dominant and typically calculated decay rate contribution, $\Gamma_\mathrm{E}$ with the total decay rate, $\Gamma_\mathrm{T}=\Gamma_\mathrm{E} + \Gamma_\mathrm{EB} + \Gamma_\mathrm{B}$. This is shown for the first three resonant modes of the cavity, where most of the variation occurs at the media interfaces and maxima/minima. Although the variation is small at a maximal difference of $0.5\%$ from $\Gamma_\mathrm{E}$ it is noticeable and becomes more significant for larger $\theta_2$ since $\Gamma_\mathrm{EB} \propto \theta_2$ and $\Gamma_\mathrm{B} \propto \theta_2^2$. This is an indicative calculation given that for $\theta_2 \neq 0$, the emitter would have to be embedded within the topological insulator medium in order to see the effect of the corrections. A more realistic treatment which involves local-field corrections should be employed to treat this in more detail, but that is beyond the scope of this paper with the central aim of demonstrating the existence of the underlying terms as well as the magnitude and structure of their effects.

\begin{figure}[htb]
	\centering
	\includegraphics[width=8.5cm, height=5.5cm]{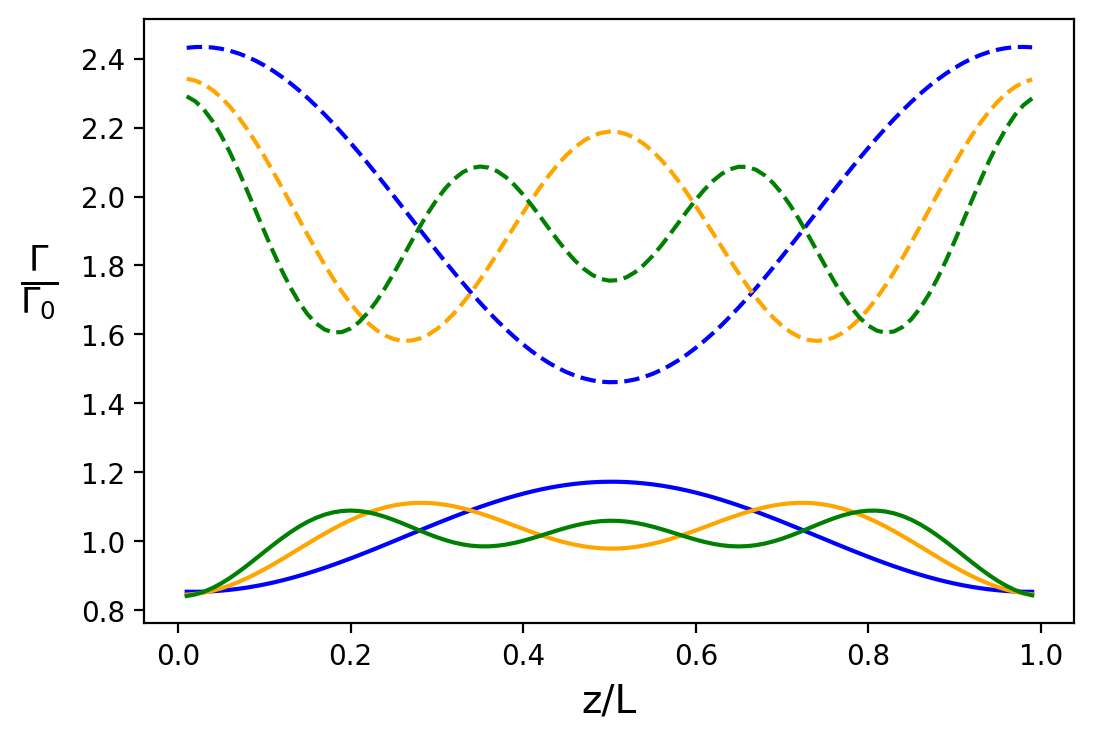}
	\caption{\label{GammaE vs GammaT (th2_400)} Position dependent decay rate normalised by $\Gamma_0$ ($\theta_1=10,\, \theta_2=400$). $\Gamma_\mathrm{E}$ (solid), $\Gamma_T$ (dashed). $\frac{\omega_\mathrm{A} L}{\pi c} = (1,2,3)$ for (blue, orange, green) lines respectively.
	}
\end{figure}

In Fig.~\ref{GammaE vs GammaT (th2_400)}, a much larger value of $\theta_2$ is chosen, which shows a considerable difference between the $\Gamma_\mathrm{E}$ and $\Gamma_T$ contributions as compared with Fig.~\ref{GammaE vs GammaT (th2_10)}. Furthermore, the nature of the position dependence in terms of the maxima and minima are completely opposite to one another, which is purely due to the inclusion of the $\Gamma_\mathrm{EB}$ and $\Gamma_\mathrm{B}$ corrections. This looks promising in terms of future study, however the largest value of $\Delta$~\eqref{Delta} experimentally realised is around 0.22 ~\cite{rivera2009short}, which corresponds to $\Delta \theta=(\theta_2-\theta_1)\approx 95$, so observing this result is not an immediate possibility while the results in Fig.~\ref{GammaE vs GammaT (th2_10)} are closer to experimental reality. Moreover, one would require careful consideration of the fact that Fermi's golden rule was employed, as this inherently makes use of the weak coupling approximation.

\begin{figure}[htb]
	\centering
	\includegraphics[width=8.5cm, height=5.5cm]{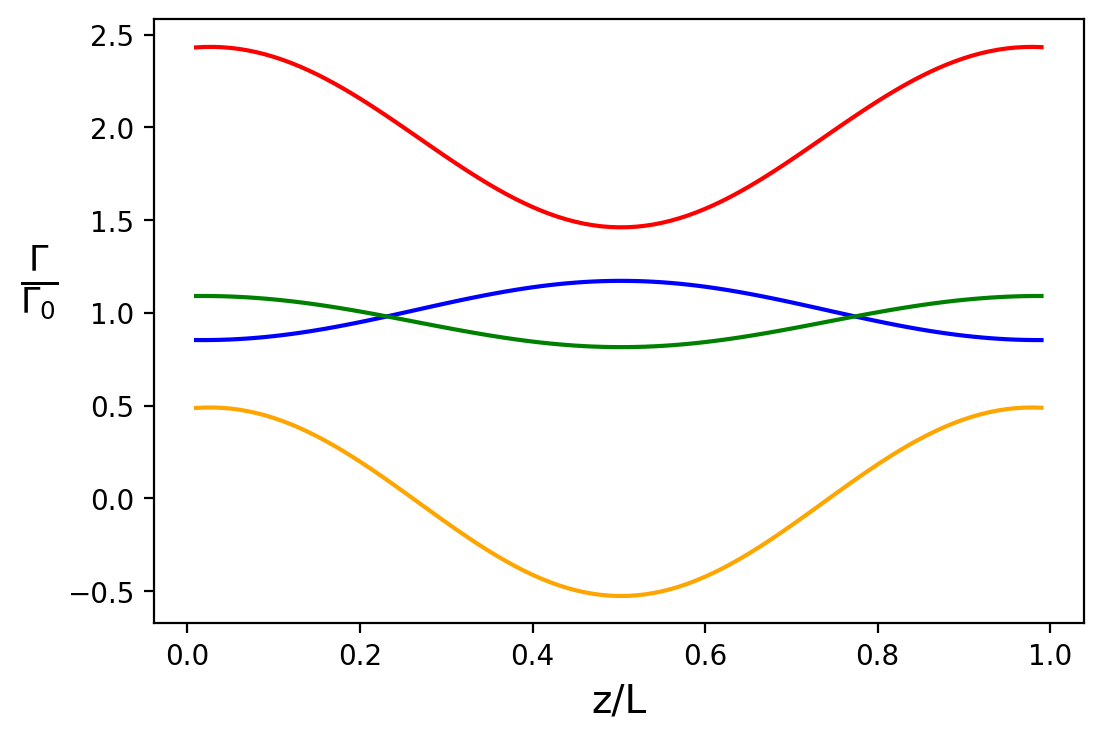}
	\caption{\label{All Gamma} Position dependent decay rate normalised by $\Gamma_0$ ($\theta_1=10,\, \theta_2=400$) for $\frac{\omega_\mathrm{A} L}{\pi c}=1$. $\Gamma_\mathrm{E}$ (blue), $\Gamma_\mathrm{EB}$ (orange), $\Gamma_\mathrm{B}$ (green), $\Gamma_T$ (red).
	}
\end{figure}
In Fig.~\ref{All Gamma} we show the individual contributions of the terms $\Gamma_\mathrm{E}$, $\Gamma_\mathrm{EB}$, and $\Gamma_\mathrm{B}$ at the first cavity mode. Due to the larger value of $\theta_2$, there is a suppression of the $\Gamma_\mathrm{EB}$ at around $z/L =0.5$, resulting in the minimum for $\Gamma_T$ where it was a maximum in $\Gamma_\mathrm{E}$. Despite this rate suppression there will never be any unphysical results such as negative decay rates as the $\Gamma_\mathrm{B}$ term grows quadratically with $\theta_2$ such that the total rate, $\Gamma_T$ is always positive. It is also important to note that the value of $\theta$ may be negative, which would physically correspond to a current in the opposite direction for the topologically protected conducting surface states of the TSB-TI media. For $\theta_2 \rightarrow -\theta_2$, this would instead flip the gain/suppression profile of $\Gamma_\mathrm{EB}$ as seen in Fig.~\ref{All Gamma} resulting in an overall enhanced profile for $\Gamma_T$ with greater gain around $z/L=0.5$. There are many combinations of $\varepsilon(\vec{r},\omega)$, $\mu(\vec{r},\omega)$, and $\theta(\vec{r},\omega)$ that could be studied, but the purpose of these results is to demonstrate the effect of the novel correction terms $\Gamma_\mathrm{EB}$ and $\Gamma_\mathrm{B}$ relative to the usual decay rate, $\Gamma_\mathrm{E}$.

\section{Discussion}

Through the canonical quantisation approach of deriving the interaction Hamiltonian, we have seen that the hidden momenta arising from the magneto-electric coupling provides an additional contribution to the interaction Hamiltonian which can have significant effects on observables such as the spontaneous decay rate in section~\ref{S5}. This approach is not unique to the non-reciprocal case of TSB-TI media, and could be applied to various other magneto-electric interactions, potentially revealing additional contributions. Similar to how macroscopic QED has been expressed for non-local and non-reciprocal media in Ref.~\cite{buhmann2012macroscopic}, it would be advantageous to generalise the canonical quantisation approach in Ref.~\cite{philbin2010canonical} for such media. This would be a daunting undertaking, but would prove fruitful in terms of having a framework for considering additional interactions with non-trivial absorbing, dispersive, inhomogeneous media, that can reveal additional interactions between a given system and the diagonalised media-field bath. 
The results derived in Eqs.~\eqref{GammaE3}, \eqref{GammaB2}, \eqref{GammaEB3} will hold for any Green's tensor, $\vec{G}$ that describes a TSB-TI medium, so further investigation can be done for different Green's tensor geometries and media properties, $\varepsilon (\vec{r},\omega)$, $\mu (\vec{r},\omega)$, $\theta (\vec{r},\omega)$. A promising next step is to incorporate local-field corrections into the analysis of $\Gamma_\mathrm{EB}$ and $\Gamma_\mathrm{B}$, since these terms only contribute when the dipole is embedded within a TSB-TI medium. Although this embedding requirement represents an experimental challenge, it also motivates the search for unique geometries and media properties in which the dipole is embedded on the surface, or stands out from the usual bulk medium.

Given the proportionality to $\theta$ and $\theta^2$ in $\Gamma_\mathrm{EB}$ and $\Gamma_\mathrm{B}$ respectively, it would be interesting to establish the limits on the range of validity of Fermi's golden rule as used in Section~\ref{S4}. In particular, the significant non-reciprocal effects on the decay rate shown in Fig.~\ref{GammaE vs GammaT (th2_400)} motivate a closer examination of the parameter range over which these effects are consistent with weak coupling. At larger values of $\theta$, stronger system--bath coupling and correlations may lead to a breakdown of the Born-Markov approximation. Among the new contributions to the decay rate, $\Gamma_\mathrm{EB}$ is particularly promising as it depends only on the off-diagonal components of $\vec{G}$ and thus provides a way to study their role in non-reciprocal response. The decay rate suppression shown in Fig.~\ref{All Gamma} provides a route towards this --- there could be potentially interesting research on minimising the total decay rate, by maximising this rate suppression for certain geometries and media properties.

\begin{acknowledgements}
The authors would like to thank Omar Jes\'{u}s Franca Santiago for early discussions around non-reciprocal media. W.M.~acknowledges financial support from  UK Engineering and Physical Sciences Research Council under grant number EP/T517896/1. N.W.~acknowledges support from the UK Engineering and Physical Sciences Research Council under grant number EP/X033015/1, as well as The Leverhulme Trust under grant number RPG-2022-335.

\end{acknowledgements}
\appendix 
\section{Constitutive Equations \& Inhomogeneous Helmholtz Equation} \label{AppendixA}
Deriving the constitutive equations for the system requires the elimination of the dynamic variables, $\vec{X}_{\omega}$ and $\vec{Y}_{\omega}$, which can be done by solving Eqs.~\eqref{XEOM} and \eqref{YEOM} with an application of the Kramers-Kronig relations in Eq.~\eqref{KKI}~\cite{philbin2010canonical}. Therefore, the frequency integrals featuring $\vec{X}_{\omega}$ and $\vec{Y}_{\omega}$ in Eqs.~\eqref{phiEOM} and \eqref{AEOM} can be solved explicitly:
\begin{align}\label{Xint}
& \int_{0}^{\infty} d\omega \, \alpha(\textbf{r}, \omega)\textbf{X}_{\omega}(\textbf{r},t) = -\varepsilon_0 \textbf{E}(\textbf{r},t) \notag \\
& + \frac{\varepsilon_0}{2\pi} \int_{0}^{\infty} d\omega \, \left[\varepsilon(\textbf{r}, \omega)\textbf{E}(\textbf{r}, \omega) e^{-i \omega t} + c.c. \right] \notag \\
& + \frac{1}{2\pi} \int_{0}^{\infty} d\omega \, \left[\alpha(\textbf{r}, \omega)\textbf{Z}_{\omega}(\textbf{r}) e^{-i \omega t} + c.c. \right]
\end{align}
and
\begin{align}\label{Yint}
& \int_{0}^{\infty} d\omega \, \beta(\textbf{r}, \omega)\textbf{Y}_{\omega}(\textbf{r},t) = \frac{1}{\mu_0} \textbf{B}(\textbf{r},t) \notag \\
& - \frac{1}{2\pi \mu_0} \int_{0}^{\infty} d\omega \, \left[\mu^{-1}(\textbf{r}, \omega)\textbf{B}(\textbf{r}, \omega) e^{-i \omega t} + c.c. \right] \notag \\
& + \frac{1}{2\pi}\int_{0}^{\infty} d\omega \, \left[\beta(\textbf{r}, \omega)\textbf{W}_{\omega}(\textbf{r}) e^{-i \omega t} + c.c. \right].
\end{align}
Here, $\vec{Z}_{\omega}(\vec{r})$ and $\vec{W}_{\omega}(\vec{r})$ are the homogeneous solutions to Eqs.~\eqref{XEOM} and \eqref{YEOM}, corresponding to oscillations of the reservoir fields with no electric or magnetic driving. Plugging Eqs.~\eqref{Xint} and \eqref{Yint} into Eqs.~\eqref{phiEOM} and \eqref{AEOM} will eliminate $\vec{X}_{\omega}$ and $\vec{Y}_{\omega}$ from the field equations of motion, giving:
\begin{align}\label{Gauss}
& \nabla \cdot \biggl( \frac{\varepsilon_0}{2\pi} \int_{0}^{\infty} d\omega \, \left[\varepsilon(\textbf{r}, \omega)\textbf{E}(\textbf{r}, \omega) e^{-i \omega t} + c.c. \right]\notag \\
& + \frac{\alpha}{\pi}Z_0^{-1}\theta(\vec{r}) \vec{B}(\vec{r},t) \biggl) 
 = -\nabla \cdot \vec{d}_{\rho}(\vec{r},t)\notag \\
 &- \frac{1}{2\pi} \nabla \cdot \int_{0}^{\infty} d\omega \, \left[\alpha(\textbf{r}, \omega)\textbf{Z}_{\omega}(\textbf{r}) e^{-i \omega t} + c.c. \right]
\end{align}
and

\begin{align}\label{Ampere}
& \nabla \times \biggl(\frac{1}{2\pi \mu_0} \int_{0}^{\infty} d\omega \, \left[\mu^{-1}(\textbf{r}, \omega)\textbf{B}(\textbf{r}, \omega) e^{-i \omega t} + c.c. \right]\notag  \\
&\qquad\qquad\qquad  - \frac{\alpha}{\pi}Z_0^{-1}\theta(\vec{r}) \vec{E}(\vec{r},t)\biggl)\notag \\
& = \partial_t \biggl( \frac{\varepsilon_0}{2\pi} \int_{0}^{\infty} d\omega \, \left[\varepsilon(\textbf{r}, \omega)\textbf{E}(\textbf{r}, \omega) e^{-i \omega t} + c.c. \right]\notag \\
& \qquad\qquad\qquad+ \frac{\alpha}{\pi}Z_0^{-1}\theta(\vec{r}) \vec{B}(\vec{r},t) \biggl)\notag \\
& + \biggl(\dot{\vec{d}}_{\rho}(\vec{r},t) + \frac{1}{2\pi} \partial_t \int_{0}^{\infty} d\omega \, \left[\alpha(\textbf{r}, \omega)\textbf{Z}_{\omega}(\textbf{r}) e^{-i \omega t} + c.c. \right] \notag\\
& + \frac{1}{2\pi} \nabla \times \int_{0}^{\infty} d\omega \, \left[\beta(\textbf{r}, \omega)\textbf{W}_{\omega}(\textbf{r}) e^{-i \omega t} + c.c. \right] \biggl).
\end{align}
Given the magneto-electric coupling in Eq.~\eqref{L_NR}, it would be expected that the usual macroscopic Maxwell's equations hold, but with the constitutive equations modified. To see this, $\vec{E}(\vec{r},t)$, $\vec{B}(\vec{r},t)$, and $\vec{d}_{\rho}(\vec{r},t)$ are written in terms of their Fourier transforms so that we have the frequency-domain equivalent of Eqs.~\eqref{Gauss} and \eqref{Ampere}:
\begin{align}\label{gaussfreq}
& \nabla \cdot \biggl( \varepsilon_0 \varepsilon(\textbf{r}, \omega)\textbf{E}(\textbf{r}, \omega) + \frac{\alpha}{\pi}Z_0^{-1}\theta(\vec{r}) \vec{B}(\vec{r},\omega) \notag \\
&+ \alpha(\textbf{r}, \omega)\textbf{Z}_{\omega}(\textbf{r}) \biggl) = -\nabla \cdot \vec{d}_{\rho}(\vec{r},\omega),
\end{align}
and
\begin{align}\label{Amperefreq}
& \nabla \times \biggl(\mu_0^{-1} \mu^{-1}(\textbf{r}, \omega)\textbf{B}(\textbf{r}, \omega) - \frac{\alpha}{\pi}Z_0^{-1}\theta(\vec{r}) \vec{E}(\vec{r},\omega) \notag \\
&- \beta(\textbf{r}, \omega)\textbf{W}_{\omega}(\textbf{r})\biggl) = -i\omega \biggl( \varepsilon_0 \varepsilon(\textbf{r}, \omega)\textbf{E}(\textbf{r}, \omega) \notag\\
&+ \frac{\alpha}{\pi}Z_0^{-1}\theta(\vec{r}) \vec{B}(\vec{r},\omega) + \alpha(\textbf{r}, \omega)\textbf{Z}_{\omega}(\textbf{r}) \biggl)  -i\omega \vec{d}_{\rho}(\vec{r},\omega).
\end{align}
Therefore, with the identifications, $\vec{P}_\mathrm{N}(\vec{r},\omega) = \alpha(\vec{r},\omega) \vec{Z}_{\omega}(\vec{r})$ and $\vec{M}_\mathrm{N}(\vec{r},\omega) = \beta(\vec{r},\omega) \vec{W}_{\omega}(\vec{r})$, we arrive at the macroscopic Maxwell's equations.
\begin{align}\label{Maxwell1}
\nabla \cdot \vec{D}(\vec{r},\omega) &= \rho_f(\vec{r},\omega), \\
\nabla \times \vec{H}(\vec{r},\omega) &= -i\omega \vec{D}(\vec{r},\omega) + \vec{J}_f(\vec{r},\omega),
\end{align}
where,
\begin{align}
 \vec{D}(\vec{r},\omega) &= \varepsilon_0 \varepsilon(\vec{r},\omega) \vec{E}(\vec{r},\omega) + \frac{\alpha}{\pi}Z_0^{-1}\theta(\vec{r})\vec{B}(\vec{r},\omega)\notag \\
& \qquad + \vec{P}_\mathrm{N}(\vec{r},\omega), \label{Dconst}\\
 \vec{H}(\vec{r},\omega) &= \mu_0^{-1} \mu^{-1}(\vec{r},\omega) \vec{B}(\vec{r},\omega) - \frac{\alpha}{\pi}Z_0^{-1}\theta(\vec{r})\vec{E}(\vec{r},\omega)\notag\\
& \qquad - \vec{M}_\mathrm{N}(\vec{r},\omega), \label{Hconst}\\
 \rho_f(\vec{r},\omega) &= \rho_d(\vec{r},\omega)=-\nabla \cdot \vec{d}_{\rho}(\vec{r},\omega), \label{rhof}\\
\vec{J}_f(\vec{r},\omega)&=\vec{J}_d(\vec{r},\omega)=-i\omega \vec{d}_{\rho}(\vec{r},\omega), \label{Jf}
\end{align}
which, except for the lack of frequency dependence in $\theta$, exactly matches the constitutive equations shown for such a non-reciprocal system in Ref.~\cite{crosse2015electromagnetic}. The frequency-dependence of this type of magneto-electric coupling factor is poorly understood and its inclusion would require a more complex initial Lagrangian, so this has been neglected to more simply showcase the effect on the interaction Hamiltonian. We can see clearly that the $\vec{Z}_{\omega}$ and $\vec{W}_{\omega}$ fields weighted by their coupling factors correspond to the noise polarisation and noise magnetisation respectively, which is reasonable as they correspond to undriven oscillations in the matter reservoirs, $\vec{X}_{\omega}$ and  $\vec{Y}_{\omega}$. Equations~\eqref{Dconst} and~\eqref{Hconst} could have equivalently been written with those noise terms expressed as free charge and current densities as in Ref.~\cite{philbin2010canonical} rather than modifications to the constitutive equations. In either case, those noise terms will correspond to fluctuations of the polariton field for the fully diagonalised system,$\,\hat{H}_0$.

Noting that $\nabla \times \vec{E} = i \omega \vec{B}$ and $\nabla \times (\theta \vec{E}) = \theta \nabla \times \vec{E} + \nabla \theta \times \vec{E}$, then the inhomogeneous Helmholtz equation can be derived from Eq.~\eqref{Amperefreq}.
\begin{align}\label{HelmholtzA}
&\nabla \times \frac{1}{\mu(\vec{r},\omega)} \nabla \times \vec{E}(\vec{r},\omega) - \frac{\omega^2}{c^2}\varepsilon(\vec{r},\omega)\vec{E}(\vec{r},\omega)\notag  \\
&- i \frac{\alpha \omega}{\pi c} \nabla \theta(\vec{r}) \times \vec{E}(\vec{r},\omega)  = \mu_0 i \omega \vec{J}_f(\vec{r},\omega)
\end{align}
which is in agreement with Ref.~\cite{crosse2015electromagnetic} given that
\begin{align}\label{HelmholtzcurrentA}
\vec{J}_f(\vec{r},\omega) &= \vec{J}_d(\vec{r},\omega) + \vec{J}_\mathrm{N}(\vec{r},\omega), \\
\vec{J}_\mathrm{N}(\vec{r},\omega) &= -i\omega \vec{P}_\mathrm{N}(\vec{r},\omega) + \nabla \times \vec{M}_\mathrm{N}(\vec{r},\omega),
\end{align}
where the dipole current density acts a source term.
\begin{align}\label{HelmholtzGreen}
&\nabla \times \frac{1}{\mu(\vec{r},\omega)} \nabla \times \vec{G}(\vec{r}, \vec{r'},\omega) - \frac{\omega^2}{c^2}\varepsilon(\vec{r},\omega)\vec{G}(\vec{r}, \vec{r'},\omega)\notag  \\
&- i \frac{\alpha \omega}{\pi c} \nabla \theta(\vec{r}) \times \vec{G}(\vec{r}, \vec{r'},\omega)  = \vec{I}_3 \delta(\vec{r} - \vec{r'})
\end{align}
The last term on the left-hand side of Eq.~\eqref{HelmholtzGreen} is directly due to the magneto-electric coupling in Eq.~\eqref{L_NR}, and it is precisely this term which breaks the Onsager reciprocity of the Green's tensor, leading to non-reciprocity of the medium. 

\section{Phenomenological Quantisation} \label{AppendixB}

Here, we follow and explicitly make use of equations from Ref.~\cite{buhmann2012macroscopic}, which outlines the general non-reciprocal, non-local phenomenological quantisation procedure. The constitutive relations for general magneto-electric coupling tensors, $\vec{\zeta}(\vec{r}, \vec{r'}, \omega)$ and $\vec{\xi}(\vec{r}, \vec{r'}, \omega)$ are given by
\begin{align}
& \hat{\vec{D}}(\vec{r}, \omega) = \varepsilon_0 \int d^3 \vec{r'}\, \vec{\varepsilon}(\vec{r}, \vec{r'}, \omega) \cdot \hat{\vec{E}}(\vec{r'},\omega) + \hat{\vec{P}}_N(\vec{r},\omega) \notag \\
& + \frac{1}{c} \int d^3 \vec{r'}\, \vec{\xi}(\vec{r}, \vec{r'}, \omega) \cdot (\hat{\vec{H}}(\vec{r'},\omega) + \hat{\vec{M}}_N(\vec{r'},\omega)) \label{DQconst} \\
& \hat{\vec{B}}(\vec{r}, \omega) = \frac{1}{c} \int d^3 \vec{r'}\, \vec{\zeta}(\vec{r}, \vec{r'}, \omega) \cdot \hat{\vec{E}}(\vec{r'},\omega) \notag \\
& + \mu_0 \int d^3 \vec{r'}\, \vec{\mu}(\vec{r}, \vec{r'}, \omega) \cdot (\hat{\vec{H}}(\vec{r'},\omega) + \hat{\vec{M}}_N(\vec{r'},\omega)), \label{BQconst}
\end{align}
\noindent which can be vastly simplified to our case in Eqs.~\eqref{Dconst} - \eqref{Hconst} with the following local, isotropic identifications
\begin{align}
&\vec{\varepsilon} (\vec{r}, \vec{r'}, \omega) =\varepsilon(\vec{r}, \omega) \vec{I}_3 \delta(\vec{r}-\vec{r'}), \label{epsilonsimp}\\
&\vec{\mu} (\vec{r}, \vec{r'}, \omega) = \mu(\vec{r}, \omega) \vec{I}_3 \delta(\vec{r}-\vec{r'}), \label{musimp}\\
& \vec{\xi} (\vec{r}, \vec{r'}, \omega) = \frac{\alpha}{\pi} \mu(\vec{r}, \omega) \theta(\vec{r}) \vec{I}_3 \delta(\vec{r}-\vec{r'}), \label{xisimp}\\
& \vec{\zeta} (\vec{r}, \vec{r'}, \omega) = \frac{\alpha}{\pi} \mu(\vec{r}, \omega) \theta(\vec{r}) \vec{I}_3 \delta(\vec{r}-\vec{r'}), \label{zetasimp}
\end{align}
\noindent where the non-reciprocity enters through the magneto-electric susceptibilities $\xi$ and $\zeta$, via our magneto-electric coupling strength, $\theta(\vec{r})$.
In order to obtain the general constitutive equations in Eqs.~\eqref{DQconst} and \eqref{BQconst}, a phenomenological quantisation procedure is carried out in Ref.~\cite{buhmann2012macroscopic} such that the induced current in the system, $\vec{J}_{in}$, is expressed generally via Ohm's law for some conductivity tensor, $\vec{Q}$
\begin{align} \label{Jinduced}
& \vec{J}_{in}(\vec{r}, \omega) = \int d^3 \vec{r'} \, \vec{Q}(\vec{r}, \vec{r'}, \omega) \cdot \vec{E}(\vec{r'},\omega) + \vec{J}_N(\vec{r}, \omega),
\end{align}
where $\vec{Q}$ is generally written as
\begin{align}
& \vec{Q}(\vec{r}, \vec{r'}, \omega) = \frac{1}{\mu_0 i \omega} \overrightarrow{\nabla}_{\vec{r}} \times \bigl(\vec{\mu}^{-1} - \vec{I}_3 \bigl) \times \overleftarrow{\nabla}_{\vec{r'}} \notag \\
& + Z_0^{-1} \overrightarrow{\nabla}_{\vec{r}} \times \bigl( \vec{\mu}^{-1} * \vec{\zeta} \bigl) + Z_0^{-1}  \bigl( \vec{\xi} * \vec{\mu}^{-1} \bigl) \times\overleftarrow{\nabla}_{\vec{r'}} \notag\\
& - i \varepsilon_0 \omega \bigl(\vec{\varepsilon} - \vec{\xi} * \vec{\mu}^{-1} * \vec{\zeta} - \vec{I}_3 \bigl), \label{Qtensor}
\end{align}
with ($*$) indicating a spatial convolution. The noise current density, $\vec{J}_N$ is quantised according to the commutator
\begin{align} \label{JNComm}
[\hat{\vec{J}}_N(\vec{r}, \omega), \hat{\vec{J}}^{\dagger}_N(\vec{r'}, \omega')] = \frac{\hbar \omega}{\pi} \mathcal{R}\mathrm{e} \vec{Q}(\vec{r}, \vec{r'}, \omega) \delta (\omega - \omega'),
\end{align}
where
\begin{align}
& \mathcal{R}\mathrm{e} \vec{Q}(\vec{r}, \vec{r'}, \omega) = \frac{1}{2} \bigl(\vec{Q}(\vec{r}, \vec{r'}, \omega) + \vec{Q}^{\dagger}(\vec{r'}, \vec{r}, \omega) \bigl) \label{ReQ}\\
& \mathcal{I}\mathrm{m} \vec{Q}(\vec{r}, \vec{r'}, \omega) = \frac{1}{2 i} \bigl(\vec{Q}(\vec{r}, \vec{r'}, \omega) - \vec{Q}^{\dagger}(\vec{r'}, \vec{r}, \omega) \bigl), \label{ImQ}
\end{align}
respectively represent the generalised real and imaginary (Hermitian and anti-Hermitian) parts of $\vec{Q}$. Given Eq.~\eqref{JNComm}, the quantised noise current density can be written as
\begin{align} \label{JNdef}
\hat{\vec{J}}_N(\vec{r},\omega) &= -i \omega \hat{\vec{P}}_N(\vec{r},\omega) + \nabla \times \hat{\vec{M}}_N(\vec{r},\omega) \notag\\
&= 
\sqrt{\frac{\hbar \omega}{\pi}} \int d^3 \vec{r'} \, \vec{R}(\vec{r}, \vec{r'}, \omega) \cdot \hat{\vec{f}}(\vec{r'}, \omega),
\end{align}
where $\vec{R}$ is defined as the positive square root of $\mathcal{R}\mathrm{e} \vec{Q}$
\begin{align} \label{Rdef}
\int d^3 \vec{s} \, \vec{R}(\vec{r}, \vec{s}, \omega) \cdot \vec{R}^{\dagger}(\vec{s}, \vec{r'}, \omega) = \mathcal{R}\mathrm{e} \vec{Q}(\vec{r}, \vec{r'}, \omega),
\end{align}
and the polariton operators, $\hat{\vec{f}}(\vec{r}, \omega)$ and $\hat{\vec{f}}^{\dagger}(\vec{r}, \omega)$ will correspondingly obey the usual bosonic canonical commutation relations
\begin{align} \label{fCCRB}
\left[\hat{f}_i(\vec{r},\omega), \hat{f}^{\dagger}_j(\vec{r'},\omega') \right] = \delta_{ij} \delta(\vec{r}-\vec{r'})\delta(\omega - \omega'),
\end{align}
while the polariton dynamics is governed by the Hamiltonian
\begin{align} \label{H0AppendixB}
\hat{H}_0 &= \hat{H}_{\mathrm{R}} + \hat{H}_{\mathrm{NR}} \notag \\
&=\int d^3 \vec{r} \int d\omega \, \hbar \omega \, \hat{\vec{f}}^{\dagger}(\vec{r}, \omega) \cdot \hat{\vec{f}}(\vec{r}, \omega).
\end{align}
It is less clear how the noise polarisation, $\hat{\vec{P}}_N$ and noise magnetisation, $\hat{\vec{M}}_N$ are expressed in terms of the polariton operators from Eq.~\eqref{JNdef}, however they can be explicitly expressed via
\begin{align} \label{PandMQuant}
\begin{bmatrix}
\hat{\vec{P}}_N \\
\hat{\vec{M}}_N
\end{bmatrix}
= \sqrt{\frac{\hbar \omega}{\pi}} \mathcal{R} * 
\begin{bmatrix}
\hat{\vec{f}}_{\mathrm{e}} \\
\hat{\vec{f}}_{\mathrm{m}}
\end{bmatrix}
\end{align}
where the $6 \times 6$ matrix $\mathcal{R}$ is the square root of
\begin{align} \label{PandMQuantexpanded}
\mathcal{R} * \mathcal{R}^{\dagger} = 
\begin{bmatrix}
\varepsilon_0 \mathcal{I}\mathrm{m} \bigl(\vec{\varepsilon} - \vec{\xi} * \vec{\mu}^{-1} * \vec{\zeta} \bigl) & \frac{\vec{\xi} * \vec{\mu}^{-1} - \vec{\zeta}^{\dagger} * (\vec{\mu}^{-1})^{\dagger}}{2 i Z_0} \\
\frac{ \vec{\mu}^{-1} * \vec{\zeta} - (\vec{\mu}^{-1})^{\dagger} * \vec{\xi}^{\dagger}}{2 i Z_0} & - \frac{\mathcal{I} \mathrm{m}(\vec{\mu}^{-1})}{\mu_0}
\end{bmatrix},
\end{align}
where the polariton operators now differ from those in Eq.~\eqref{JNdef} as they are split into electric excitations,  $\hat{\vec{f}}_{\mathrm{e}}(\vec{r}, \omega)$, and  magnetic excitations, $\hat{\vec{f}}_{\mathrm{m}}(\vec{r}, \omega)$, which obey
\begin{align} \label{fCCRlambda}
\left[\hat{f}_{i,\lambda}(\vec{r},\omega), \hat{f}^{\dagger}_{j, \lambda'}(\vec{r'},\omega') \right] = \delta_{ij} \delta_{\lambda \lambda'} \delta(\vec{r}-\vec{r'})\delta(\omega - \omega'),
\end{align}
where $\lambda=(e,m)$ distinguishes the type of excitation, and the dynamics of these operators is governed by
\begin{align} \label{H0AppendixBlambda}
\hat{H}_0 =\sum_{\lambda}\int d^3 \vec{r} \int d\omega \, \hbar \omega \, \hat{\vec{f}}^{\dagger}_{\lambda}(\vec{r}, \omega) \cdot \hat{\vec{f}}_{\lambda}(\vec{r}, \omega).
\end{align}
This is the clearest way to express the noise polarisation and magnetisation explicitly in terms of the polariton operators, and it is possible, albeit complicated, to then relate $\hat{\vec{f}}_{\mathrm{e}}(\vec{r}, \omega)$ and $\hat{\vec{f}}_{\mathrm{m}}(\vec{r}, \omega)$ to the more general polariton operators, $\hat{\vec{f}}(\vec{r}, \omega)$, featuring in Eq.~\eqref{JNdef}.

Following from Eq.~\eqref{Jinduced}, the inhomogeneous Helmholtz equation can be written as
\begin{align}
& \left[\nabla \times \nabla \times \vec{E}(\vec{r},\omega) - \frac{\omega^2}{c^2} \vec{E}(\vec{r},\omega)\right] \notag\\
& - \mu_0 i \omega \int d^3 \vec{r'} \, \vec{Q}(\vec{r}, \vec{r'},\omega) \cdot \vec{E}(\vec{r'},\omega) = \mu_0 i \omega \vec{J}_N(\vec{r},\omega),
\end{align}
which upon applying the identifications in Eqs.~\eqref{epsilonsimp} - \eqref{zetasimp} will simplify to the inhomogeneous Helmholtz equation for our case in Eq.~\eqref{Helmholtz} (without the source current density). After quantising the system with Eq.~\eqref{JNdef}, then the quantised electric field can be written as the particular solution to the inhomogeneous Helmholtz equation
\begin{align}
& \hat{\vec{E}}(\vec{r},\omega) = \mu_0 i \omega \int d^3 \vec{r'} \, \vec{G}(\vec{r}, \vec{r'}, \omega) \cdot \hat{\vec{J}}_N(\vec{r'},\omega) \notag \\
&= \mu_0 i \omega \sqrt{\frac{\hbar \omega}{\pi}} \int d^3 \vec{r'} \int d^3 \vec{s} \, \vec{G}(\vec{r}, \vec{r'}, \omega) \cdot \vec{R}(\vec{r'}, \vec{s}, \omega) \cdot \hat{\vec{f}}(\vec{s},\omega),
\end{align}
where $\vec{G}$ is the Green's tensor of the inhomogeneous Helmholtz equation in Eq.~\eqref{HelmholtzA} and the quantised magnetic field can be correspondingly written using $i \omega \hat{\vec{B}}(\vec{r},\omega) = \nabla \times \hat{\vec{E}}(\vec{r},\omega)$.

\bibliography{WM.bib,RB.bib}

\end{document}